\documentclass[mnsc,nonblindrev]{informs3_SSRN}

\OneAndAHalfSpacedXI % current default line spacing
\usepackage{natbib}
\def\bibfont{\small}%
\usepackage{endnotes}
\let\footnote=\endnote

\TheoremsNumberedThrough     % Preferred (Theorem 1, Lemma 1, Theorem 2)
\ECRepeatTheorems

\EquationsNumberedThrough    % Default: (1), (2), ...
\MANUSCRIPTNO{OPRE-0000-0000.00}

\usepackage{color}
\usepackage{multimedia}
\usepackage{multicol}
\usepackage{bbm}
\usepackage{graphics}
\usepackage{graphicx}
\usepackage{booktabs}
\usepackage{romannum}

\usepackage{grffile}
\usepackage{float}
\usepackage{url}
\usepackage{mathabx}
\usepackage{supertabular}
\usepackage{rotating}
\usepackage{pdflscape}
\usepackage{verbatim}
\usepackage{listings}
\usepackage{xcolor}
\usepackage{multicol}
\usepackage[shortlabels]{enumitem}
\usepackage{caption,subcaption}
\usepackage{epstopdf}
\usepackage{algorithm}
\usepackage{algorithmic}
\usepackage{enumitem}
\usepackage{pifont}

\newcommand{\dd}{\mathrm{d}}
\newcommand{\one}{\mathbbm{1}}
\newcommand{\E}{\mathbb{E}}
\newcommand{\p}{\mathbb{P}}

\newcommand{\bm}[1]{\boldsymbol{#1}}

\graphicspath{{Figures/}}

\begin{document}
	%%%%%%%%%%%%%%%%
	\pagenumbering{arabic}
	%\setcounter{page}{1}
	%\DOI{10.1287/xxxx.0000.0000}%
	
	\RUNAUTHOR{Dai et al.}
	
	\RUNTITLE{Data-Driven Merton's Strategies  via Policy Randomization}
	
	% \TITLE{Bundling Information Goods of Decreasing Value}
	\TITLE{Data-Driven Merton's Strategies via Policy Randomization}
	
	\ARTICLEAUTHORS{%
		\AUTHOR{Min Dai\footnote{Dai acknowledges the supports of Hong Kong GRF (15213422, 15217123), The Hong Kong Polytechnic University Research Grants (P0039114, P0042456, P0042708, and P0045342), and NSFC (12071333).}}
		\AFF{Department of Applied Mathematics and School of Accounting and Finance, The Hong Kong Polytechnic University, Hung Hom, Hong Kong, Kowloon, \EMAIL{mindai@polyu.edu.hk}} %, \URL{}}
	\AUTHOR{Yuchao Dong}
	\AFF{School of Mathematical Sciences\\ Tongji University, Shanghai, China, Shanghai 200092, \EMAIL{ycdong@tongji.edu.cn}}
	\AUTHOR{Yanwei Jia}
	\AFF{Department of Systems Engineering and Engineering Management, The Chinese University of Hong Kong, Shatin, Hong Kong, New Territories, \EMAIL{yanweijia@cuhk.edu.hk}}
	\AUTHOR{Xun Yu Zhou}
	\AFF{Department of Industrial Engineering and Operations Research \& Data Science Institute, Columbia University, New York, USA, NY 10027, \EMAIL{xz2574@columbia.edu}}
} % end of the block

\ABSTRACT{%
	We study Merton's expected utility maximization problem in an incomplete market, characterized by a factor process in addition to the stock price process, where the functional forms of the model primitives are unknown. The agent under consideration is a price taker who has access only to the stock and factor value processes and the instantaneous volatility. We propose an auxiliary problem in which the agent can invoke policy randomization according to a specific class of Gaussian distributions, and prove that the mean of its optimal Gaussian policy solves the original Merton problem. With randomized policies, we are in the realm of continuous-time  reinforcement learning (RL) recently developed in \citet{wang2018exploration} and \citet{jia2021policy,jia2021pgac,jia2023q}, enabling us to solve the auxiliary problem in a data-driven way without having to estimate the model primitives. Specifically, we establish a policy improvement theorem based on which we design both online and offline  actor-critic  RL algorithms for learning Merton's strategies. A key insight from this study is that RL in general and policy randomization in particular are useful beyond the purpose for exploration---they can be employed as a technical tool to solve a problem that cannot be otherwise solved by mere deterministic policies.   Finally, we present simulation and empirical studies in a stochastic volatility environment to demonstrate the decisive outperformance of the devised RL algorithms in comparison to the conventional model-based, plug-in method.
}%

%\KEYWORDS{deterministic inventory theory; infinite linear programming duality;
	%  existence of optimal policies; semi-Markov decision process; cyclic schedule}

%\HISTORY{This paper was first submitted on 8 August 2020.}

\maketitle
%%%%%%%%%%%%%%%%%%%%%%%%%%%%%%%%%%%%%%%%%%%%%%%%%%%%%%%%%%%%%%%%%%%%%%

% Samples of sectioning (and labeling) in MNSC
% NOTE: (1) \section and \subsection do NOT end with a period
%       (2) \subsubsection and lower need end punctuation
%       (3) capitalization is as shown (title style).
%
%\section{Introduction.}\label{intro} %%1.
%\subsection{Duality and the Classical EOQ Problem.}\label{class-EOQ} %% 1.1.
%\subsection{Outline.}\label{outline1} %% 1.2.
%\subsubsection{Cyclic Schedules for the General Deterministic SMDP.}
%  \label{cyclic-schedules} %% 1.2.1
%\section{Problem Description.}\label{problemdescription} %% 2.

% Text of your paper here

\section{Introduction}
\label{intro}

Merton's expected utility maximization model \citep{merton1969lifetime} and its subsequent rich variants are central to continuous-time finance.
%Dynamic asset allocation has been a central topic in the modern financial industry, which aims to construct and adjust portfolios dynamically to incorporate investors' risk preferences into investment decisions. An individual's risk preference, studied in decision science and microeconomics, is usually characterized by a utility function. Theoretical analysis of investors' utility-based dynamic optimal investment decisions (under a continuous-time framework) can be traced back to \citet{merton1969lifetime}, followed and extended by extensive literature. However, there is still a lack of practical implementation and application of Merton's strategies. In particular, owing to recent innovations in machine learning, there is a growing interest in integrating mass financial data to assist investors in allocating wisely. An end-to-end approach that maps data to the investment decision is the foundation of computer-assisted portfolio management or robotic financial advising systems and has been a key focus of many FinTech startups.
The traditional paradigm for applying the Merton models to practice follows the so-called
``separation principle" (separation between estimation and optimization), also known as the ``plug-in" method. Starting from a baseline stock price model---be it the simplest Black-Scholes,  a stochastic volatility model, or a jump-diffusion---an econometrician estimates the model parameters/primitives
from historical data using statistical or machine learning methods and then passes on to an (optimization) theorist. The latter plugs in the estimated values to the resulting stochastic control problem and solves it (rarely) analytically or (commonly) numerically via solving Pontryagin's maximum principle conditions or  Hamilton-Jacobi-Bellman (HJB) partial differential equations (PDEs). The endeavors of the econometrician and the theorist are thus {\it separated}: The former deals with estimation only, and the latter takes the estimated model as given and focuses on optimization.

{\it Had} an infinite amount of data been available, this division of labor might %actually
work: appropriate statistical/econometric methods would ensure that the model is validated with reasonable accuracy and that the primitives are estimated as precisely as possible. However, in financial markets, it is well documented that accurately estimating certain parameters---predominantly the expected return---requires far more data than the available historical record \citep{merton1980estimating, luenbergerinvestment}. Even worse, financial markets are likely non-stationary, violating the stationarity assumption underlying many econometric methods.  Furthermore, Merton's strategies, if computable, are typically highly sensitive to model primitives.
Consequently, estimation errors can propagate and amplify through the optimization step, often rendering the resulting strategies impractical.

In contrast, the modern reinforcement learning (RL) paradigm adopts a {\it conceptually} and fundamentally different approach.\footnote{RL has been predominantly studied in discrete-time Markov decision processes (MDPs); see, e.g., \cite{sutton2011reinforcement} for a comprehensive treatment.} It still begins with a basic structural model underlying the data-generating process (e.g., a Markov chain or a diffusion process), but it does not assume the model parameters to be given and known, {\it nor does it  attempt to estimate them}. Instead, RL seeks to learn optimal policies or strategies {\it directly} by parameterizing a policy and iteratively updating (learning) its parameters to improve performance until optimality or near-optimality is achieved. This process typically involves three key steps: 1) strategically exploring the unknown environment (e.g., a market) by trial-and-error:  randomly experimenting with different choices according to some carefully designed probability distribution (called a {\it randomized} or {\it stochastic} policy) and observing the responses (called reward or reinforcement signals) from the environment; 2) learning the value function of that randomized policy based on the reward signals; and 3) improving the randomized policy based on the learned value function. These steps are called respectively {\it exploration, policy evaluation}, and {\it policy improvement}, and the resulting algorithms are referred to as the
{\it actor} (policy)-{\it critic} (value function) type in RL. So RL is end-to-end, model-free and data-driven: It maps data to decision policies, skipping the middle step of estimating model primitives.\footnote{Throughout this paper, by ``model-free" we mean that we do not have access to  the model primitives, although---as mentioned earlier---we do have a basic structural model  such as a diffusion process, as in this paper. By ``data-driven" we mean that policies are learned by observable/computable data---both exogenous and endogenous---such as stock prices and volatility  processes.}

The primary motivation for employing randomized policies in RL is exploration: randomization expands the search space and enables the observation of counterfactual outcomes of alternative actions that deterministic policies would never try, thereby facilitating a better understanding of the interaction between actions and the environment. However, this motivation appears irrelevant in the classical Merton problem for a small investor (i.e., a price taker). Since stock prices are exogenous to such an investor, the returns of any portfolio can be computed without actual execution (e.g., via a “paper portfolio”). Hence, there seems to be no informational advantage in adopting randomized policies or the RL framework for small investors.

This paper challenges this conventional wisdom and demonstrates that RL can nonetheless be employed to solve Merton’s problem efficiently and effectively in a model-free, data-driven manner even for small investors. We construct an auxiliary problem that relaxes the original Merton problem by allowing a certain class of randomized policies---specifically, Gaussian policies with a particular variance structure. This formulation is inspired by the stochastic relaxed control framework introduced by \citet{wang2018exploration} for continuous-time RL. We prove that the mean of the optimal Gaussian policy for the auxiliary problem coincides with the optimal policy of the original Merton problem, thereby justifying the auxiliary formulation and the use of randomized policies. More importantly, once randomized policies are incorporated, the problem falls naturally within the RL paradigm, enabling us to develop RL algorithms based on the general theory and methods established in \citet{jia2021policy,jia2021pgac,jia2023q}. In particular, we design an algorithm tailored to the Merton problem with power utility functions by leveraging their homothetic properties to enhance computational efficiency. We prove convergence of the proposed algorithm in the Black-Scholes market, achieving the ``optimal” convergence rate typical in the literature. Notably, we show that randomized policies are necessary for our algorithms to function: the algorithms do not update deterministic policies, as demonstrated by analyzing the impact of the temperature parameter controlling the degree of randomization. When randomized policies degenerate to point masses (deterministic policies), the variance of the reinforcement signal becomes prohibitively large, preventing effective policy improvement.

Using a stochastic volatility model studied by \citet{liu2007portfolio}, we further illustrate why RL outperforms the plug-in method even when the model class is correctly specified but its parameters are unknown. Beyond the statistical challenges in estimating model primitives, a subtle but critical difficulty arises with the plug-in approach: the forms of model-based optimal policies can depend drastically on the constellations of model parameters, some of which may be practically unreasonable or infeasible. Statistical estimation methods typically do not incorporate such parameter constraints, focusing solely on fitting the model without considering the implications for subsequent decision-making. As a result, estimation errors can propagate into significant policy errors. In contrast, the RL approach begins with a reasonable policy class and iteratively improves policies within that class, thereby effectively circumventing this issue.

Finally, we present simulation and empirical studies comparing the performance of our RL algorithms against the classical plug-in method, a na\"ive buy-and-hold strategy and another data-driven method called empirical risk minimization (ERM). Our results demonstrate that the RL methods consistently exhibit superior robustness and overall performance.

\subsection*{Related Literature}
The original Merton problem \citep{merton1969lifetime} is formulated under a
Black-Scholes market setting.
Subsequent studies involve more general and richer market models, e.g., ones in which instantaneous mean return and/or volatility are driven by additional random sources. The corresponding Merton problem has been studied in, to name but a few, \citet{wachter2002portfolio,chacko2005dynamic,liu2007portfolio}. The literature on the Merton problem has primarily taken the perspective of an economic agent who,  having access to a correct  market model, focuses on solving the portfolio selection problem and provides insights into how different market conditions affect optimal portfolio choices and asset prices. There are papers addressing the agent's incomplete information on the expected stock returns, and they either assume that the agent conducts Bayesian learning (e.g., \citealt{gennotte1986optimal,pastor2000portfolio,cvitanic2006dynamic,andrei2015investor}), or take a robust control approach to consider the worst scenario among a model class (e.g., \citealt{hansen2001robust,maenhout2004robust,hansen2006robust}). However, the former Bayesian approach has been restricted to simple models to keep the Bayesian updating tractable for analysis and computation, and the latter robust approach crucially relies on specifying a class of models while the model uncertainty is not endogenously determined.%\footnote{\citet{epstein2007learning} discuss how to incorporate learning into model ambiguity, but the analysis is not tractable for complex models if the Bayesian posterior is not explicitly available.} 
To our best knowledge, no paper systematically studies the Merton problem for an investor with minimum knowledge about a ``model''  who learns optimal choice in both offline and online settings.
The present paper aims to fill this void---it tackles the problem in an incomplete market by developing interpretable and efficient algorithms that learn the optimal
policy without knowing or trying to estimate the market specifications.

This paper relates to a strand of literature on machine learning for financial decision making, especially its applications to dynamic portfolio choice. \citet{gao2000algorithm} and \citet{jin2016portfolio} formulate a Merton problem as a discrete-time MDP that allows only a finite number of decisions. They apply Q-learning algorithms with portfolio return as a reward with/without adjusting for risk. In contrast, we consider continuous-time and continuous state-action spaces to reflect more realistic trading patterns including high-frequency transactions and allocation of an arbitrary percentage of total wealth to risky assets.\footnote{For ease of presentation, in this paper, we consider a market with only one risky asset (e.g., a market index fund), but our method can be readily generalized to multiple assets.} There have also been attempts to employ deep neural networks to solve MDPs with continuous state-action spaces or stochastic control problems; see, e.g.,  \citet{han2016deep,bachouch2021deep}, and \citet{duarte2024machine}. However, these papers assume the models are completely known and apply neural networks only as a computational tool to solve the respective optimization problems. As such, their approaches are alternatives to the traditional simulation or PDE-based numerical methods, instead of providing end-to-end solutions that map data to decisions. On the other hand, there are  works that directly learn {\it deterministic} trading policies  via the so-called ``empirical risk minimization (ERM)"; see,  for example, \citet{guijarro2021deep} for the one-period mean-variance model and \citet{buehler2019deep} for dynamic hedging. However, ERM can only do offline learning as it inherently requires the data in the whole time horizon while our method permits both online and offline learning. Moreover,  \citet{reppen2023deep} demonstrate that ERM tends to perform poorly with limited data sets and exhibits desired convergence only with sufficiently large data sets. In the present paper, we will also show (in Appendix B) that our RL algorithms
perform better than ERM when the sample size is small.

In recent years, there has been an upsurge of interest in continuous-time RL with continuous state and action spaces, not only because many practical problems are continuous time by nature (e.g., autonomous driving, robot navigation, and ultra-high-frequency trading)  but also because more analytical tools are available in the continuous setting for developing a rigorous theory. Works by \citet{wang2018exploration,jia2021policy,jia2021pgac,jia2023q} lay the theoretical foundation for the formulation and algorithm design for continuous-time RL. A central underpinning of this series of research is the {\it martingality}: the learning/updating of the parameters of both the actor and the critic is guided by maintaining the martingality for various stochastic processes. Applications of these general results include, to name just a few, \citet{wang2019continuous} for continuous-time mean-variance pre-committed portfolio choice, \citet{ddj2020calibrating} for mean-variance equilibrium policies, \citet{wang2023reinforcement} for liquidation and execution, and \citet{guo2020entropy} for mean-field games. However, they have been largely restricted to the class of linear-quadratic problems. The present paper is the first to apply continuous-time RL to  utility-based portfolio selection.

The rest of the paper is organized as follows. Section \ref{setup} introduces the model-free Merton problem as well as an auxiliary problem with randomized policies.  In Section \ref{sec:theory analysis} we conduct a theoretical analysis including the connection between the original problem and the auxiliary one, based on which we present both online and offline RL algorithms to learn the optimal policy. We also prove convergence of the offline algorithm for the special Black-Scholes setting. Next, we use a class of stochastic volatility models to illustrate the benefits of the RL methods in Section \ref{sec:sv}. A simulation study and an empirical analysis are presented in Section \ref{numerical}. Finally, Section \ref{sec:concl} concludes. All the proofs and additional results/discussions are placed in the appendix.%\footnote{The code to reproduce the numerical results in this paper is available at \url{https://www.dropbox.com/scl/fo/onrln1ggs3vl46aclgno9/AMFr2VV2U1mQRNSScVyi300?rlkey=t55jyru3y9u9xttzncpbsorpy&st=fj6csbtu&dl=0}.}

%How to incentivize exploration through a proper regularizer in reinforcement learning algorithms in a reasonable way has long been a well-known problem. There are many ad-hoc or exogenous ways to achieve this by some naive rules, for example, entropy regularizer with a finely tuned temperature parameter.
\section{Problem Formulation}\label{setup}
Throughout this paper, % with a slight abuse of notation, we use either $Z$ or $Z_t$ to refer to a stochastic process $Z:=\{Z_t\}_{t\in[0,T]}$, while $Z_t$ may also refer to the value of the process at time $t$ if it is clear from the context. We use $f(\cdot)$ or $f$ to denote a function, and $f(x)$ to denote the value of the function $f$ at $x$. 
for a function $f$ with  arguments $(t,w,x)$, we use $\frac{\partial f}{\partial t}$, $f_w,f_x,f_{ww},f_{wx},f_{xx}$ to denote its first- and second-order partial derivatives with respect to the arguments.
We use bold-faced $\bm\pi$ to denote various probability-density-function valued portfolio controls or policies, and $\pi \approx 3.14$ and $e\approx 2.72$ to denote the respective mathematical constants. For a probability density function $\bm\pi$ on $\mathbb{R}$, we denote its mean and variance by $\operatorname{Mean}(\bm\pi) = \int_{\mathbb{R}} a\bm\pi(a)\dd a$ and $\operatorname{Var}(\bm\pi) = \int_{\mathbb{R}} a^2 \bm\pi(a)\dd a - \operatorname{Mean}(\bm\pi)^2$, respectively. Finally, we denote by $\mathcal{N}(a,b^2)$ the density function of a normal distribution with mean $a$ and variance $b^2$, and $\mathcal N(a,0)$ by Dirac mass at point $a$.

\subsection{Market Environment and Investment Objective}

There are two assets available for investment in a market: a risk-free asset (bond) with a constant interest rate $r$ and a stock (or market index). The stock price process is observable, whose dynamic is governed by the following stochastic differential equation (SDE):
\begin{equation}
	\label{eq:model stock}
	\frac{\dd S_t}{S_t}=\mu(t,X_t)\dd t+\sigma(t,X_t)\dd B_t,\ S_0=s_0,
\end{equation}
where $B$ is a scalar-valued Brownian motion, and the instantaneous return rate process $\mu_t\equiv \mu(t,X_t)$  and  volatility process $\sigma_t\equiv \sigma(t,X_t)$ both depend on another observable stochastic market factor process $X$. We assume that $X$ follows SDE:
\begin{equation}
	\label{eq:model factor}
	\dd X_t =m(t,X_t)\dd t+\nu(t,X_t)[\rho \dd B_t +\sqrt{1-\rho^2}\dd \tilde B_t],\ X_0 = x_0,
\end{equation}
where $\tilde B$ is another (scalar-valued) Brownian motion independent of $B$, and $\rho\in(-1,1)$ is a constant that determines the correlation between the stock return and the change in the market factor. So the market is in general incomplete. We only consider the Markovian model, i.e., $ \mu(\cdot,\cdot)$, $ \sigma(\cdot,\cdot)$, $ m(\cdot,\cdot)$, and $ \nu(\cdot,\cdot)$ are deterministic and continuous functions of $t$ and $x$ such that equations \eqref{eq:model stock}-\eqref{eq:model factor} have a unique weak solution. %However, the exact forms of these functions, along with the value of the constant $\rho$, are {\it unknown} to the investor. %We use $\mu_t, \sigma_t, \sigma_t, \nu_t$ to denote the process $ \mu(t,X_t)$, $ \sigma(t,X_t)$, $ m(t,X_t)$, $ \nu(t,X_t)$ wherever no confusion is caused.
This market setup is similar to that of \citet{dai2018dynamic}, which covers many popular and incomplete market models as special cases, e.g., \citet{wachter2002portfolio}, \citet{liu2007portfolio}, and \citet{chacko2005dynamic} among others.

A (small) investor's actions are modeled as a scalar-valued adapted process $a=\{a_t\}_{t\in[0,T)}$, with $a_t$ representing the fraction of total wealth invested in the stock at time $t$. The corresponding
self-financing wealth process $W^a$ then follows the SDE:
\begin{equation}\label{classical_wealth}
	\frac{\dd W_t^a}{W^a_t}= a_t \frac{\dd S_t}{S_t} + (1-a_t)r\dd t =[r+(\mu(t,X_t)-r)a_t]\dd t+\sigma(t,X_t) a_t\dd B_t,\ \ W_0^a = w_0.
\end{equation}
Note that the solvency constraint $W_t^a\geq 0$ a.s., for all $t\in [0,T]$, is automatically satisfied for any square-integrable $a$.
The Merton investment problem is to choose $a$ to maximize the following expected utility of the terminal wealth:
\begin{equation}
	\label{objective_functional_classical}
	\mathbb E\left[ U(W_T^a)\right],
\end{equation}
where $W_T^a$ is defined by \eqref{eq:model factor}-\eqref{classical_wealth} and $U(\cdot)$ is a utility function. We focus on the constant relative risk aversion (CRRA) utility function in the main body of this paper, i.e., $U(w) = \frac{w^{1-\gamma} - 1}{1-\gamma}$, where $1\neq\gamma > 0$ is the relative risk aversion coefficient.\footnote{When $\gamma = 1$, the CRRA utility function becomes the logarithm function $U(w) = \log w$. A problem with log utility can be regarded as a special case of  the mean-variance problem for log returns in \cite{ddj2020calibrating} and \citet{jiang2022reinforcement}. Hence we restrict our attention to the case of $\gamma \neq 1$ in this paper.} %We will discuss the constant absolute risk aversion (CARA) utility function in Appendix \ref{sect: cara}.

%The common approach to solve the problem \eqref{objective_functional_classical} is via dynamic programming and Hamilton-Jacobi-Bellman (HJB) equation, which requires defining a family of problems starting from different initial time and states. That is,
%\begin{equation}
%\label{eq:classical optimal value function def}
%V(t,x,w) = \max_{\{a_s\}_{s\in [t,T)}} \mathbb E\left[ e^{-\beta (T - t)} U(W_T^{a}) \Big| X_t = x, W_t^{a} = w \right] .
%\end{equation}
%However, the conventional approach relies on solving a non-linear partial differential equation (PDE) that depends on the knowledge about the market.

\subsection{Agent's Knowledge and Randomized Choices}
\label{subsec:incentive}

The classical Merton problem is model-based, namely, all the model primitives are assumed to be known and given, %\change{my edit -- The classical Merton problem is studied for an agent who has formed the correct belief about the market environment, namely, all the model primitives are assumed to be known and given,  -- please check and decide}
and the problem is typically solved by dynamic programming and HJB equations, leading to a {\it deterministic} optimal (feedback) policy. %Hence, the classical problem describes an oracle agent's optimal investment behavior: The oracle agent is omniscient about the market environment to form a rational expectation of a ``model'' for the stock prices that is consistent with the actual data; but the agent is not perfect foresighted and cannot peek into the future. Being an oracle is an ideal assumption, and literature has investigated this problem under incomplete information for Bayesian agents who are endowed with certain prior about a class of models and update their beliefs about each model gradually as they observe more data.

In this paper, however, we consider an agent who does not have knowledge about the market environment up to the diffusion structure presented in the previous subsection and is unable to form a proper prior on each model within the family specified in \eqref{eq:model stock} and \eqref{eq:model factor} or unable to do Bayesian update of beliefs on each model. This setting is motivated by the difficulty in computing Bayesian posterior on general functional spaces, as well as the difficulty in specifying priors.\footnote{Even if each model is indexed by a finite-dimensional vector, it is already challenging in posterior computation beyond the conjugate family (cf. \citet{green2015bayesian}).} The agent encounters multiple investment episodes, each with horizon $T$. Within each episode, the time is indexed by $t\in [0,T]$, and at time $t$, the past trajectories of the stock-factor value process  and  the wealth-portfolio process up to time $t$ within the current episode can be observed. For simplicity, we assume the relative risk aversion coefficient $\gamma$ is known to the agent. Finally, for the particular approach employed in this paper we need to assume that the volatility process $G_t=\sigma(t,X_t)^2$ is observable.
This assumption is premised upon the well-documented results that the volatility
may be approximated accurately by VIX, option data, or high-frequency observation of stock returns.\footnote{For example, instantaneous variance can be calculated accurately based on the realized variance with high-frequency observations \citep{barndorff2002estimating, hansen2006realized}. Alternatively, it is possible to use the derivative price on realized variance \citep{carr2005pricing} as a proxy for the instantaneous variance, such as VIX for S\&P 500 index.}

So the agent's task is to solve the Merton problem in a data-driven way, where data include only stock-factor-volatility processes, the agent's own wealth process under any given portfolio, and the risk aversion parameter, without knowledge of the forms of market coefficients $\mu(\cdot,\cdot),\sigma(\cdot,\cdot),m(\cdot,\cdot),\nu(\cdot,\cdot)$. This knowledge/information structure more accurately reflects an actual investor's knowledge rather than a hypothetical omniscient agent. Moreover, in reality, when faced with an unknown environment, humans tend to use trial-and-error to test various strategies (i.e. engage randomized policies) and learn from experience.%\footnote{It is interesting to note that taking randomized decisions is often observed in behavioral experiments \citep{agranov2017stochastic} and  considered as an integral part of human behaviors \citep{mattsson2002probabilistic,swait2013probabilistic}. Randomized policies are popular in analyzing (dynamic) discrete choices \citep{hotz1993conditional}, and our setting here  is a natural extension to accommodate continuum choices.}

{Both the knowledge structure and the employment of randomized policies are} prevalent in the general RL literature. A distinctive feature of RL compared with standard optimization or statistics is that  ``data" can be {\it endogenous} and, hence, also part of the solutions. It is generally acknowledged that a policy in RL has two objectives: to learn the environment {\it relevant} to the optimization objective and to improve performance. The former is the demand for exploration while the latter for exploitation. The essence of RL is to strike the best exploration-exploitation balance, which is usually achieved by randomizing decisions, i.e., extending the policy space to include stochastic or randomized policies (or mixed strategies in game theory). It is randomization that generates endogenous data for learning. %However, randomization goes beyond the need for exploration. Indeed, it is technically  necessary and essential for designing learning algorithms, as we will show in this paper in the context of investment problems.

%Before theoretical analysis, we must first establish an analytical framework.
Following \cite{wang2018exploration}, we now reformulate the Merton problem with randomized policies. An investor chooses her time-$t$ action (portfolio) by sampling from a probability distribution $\bm\pi_t$, where $\{\bm\pi_t\}_{t\in[0,T]}=:\bm\pi$ is a distribution-valued process called a stochastic or exploratory {\it control}. The resulting {\it exploratory dynamic} of the wealth process is described by
\begin{equation}
	\label{controlled_system}
	\frac{\dd W^{\bm\pi}_t}{W^{\bm\pi}_t} = \left[r+(\mu(t,X_t)-r)\operatorname{Mean}(\bm\pi_t)\right]\dd t +\sigma(t,X_t) \left[ \operatorname{Mean}(\bm\pi_t) \dd B_t+\sqrt{ \operatorname{Var}(\bm\pi_t) }\dd \bar{B}_t\right],\ W^{\bm\pi}_0 = w_0,
\end{equation}
where $\bar B$ is another Brownian motion that is independent of both $B$ and $\tilde B$, characterizing the additional noises introduced into the wealth process due to randomization.
Intuitively, \eqref{controlled_system} is the limit of equations where actions are sampled from the randomized policy $\boldsymbol{\pi}$ at discrete times. %limit process of the discrete time models as the time step $\Delta t$ goes to zero.
%It is motivated by central limit theorem. Thus,
%An extra Brownian motion $\bar B$ is introduced and only the mean and variance of $\pi$ appear in \eqref{controlled_system}.
The derivation of \eqref{controlled_system} is analogous to that in \cite{ddj2020calibrating} and an informal explanation is provided in Appendix \ref{appendix:motivation}. {A rigorous proof of how \eqref{controlled_system} describes the wealth process under the random portfolio choices is presented in \citet{jia2025accuracy}.}

%Now maximizing the objective function \eqref{objective_functional_classical} with $W_T^{\bm\pi}$ is defined by the law of motion \eqref{controlled_system} and \eqref{eq:model factor} over a set of stochastic policies becomes a valid theoretical question (where the admissible policies will be defined later). However,

Finally, we reiterate the  important point about the need and interpretation of randomized decisions in the particular Merton problem {\it with a small investor} that differs from the general RL.
%Besides the behavioral or irrational reasons for adopting randomized decisions,
%Does exploration--exploitation tradeoff provide a reason for randomization in our particular portfolio choice problem? The answer is {\it no}!
The rationale of using randomization for exploration is to learn how the (unknown) environment reacts to a greater number of different decisions. This rationale is only valid when such a reaction is unknown a priori. For example, one will not observe the return of a slot machine (the counterfactual) unless actually playing it. However, in the setting of this paper with a small investor, how the environment (market) reacts to the agent's decision (portfolio choice) can be {\it deduced}, as shown in the first equation  in \eqref{classical_wealth}. %\footnote{Such assumption applies mainly to small retailer investors but does not apply to large institutional investors.}
Hence, observing the counterfactual returns of alternative portfolios is possible without having  to actually execute those portfolios to gain information about the market. Therefore, the primary motivation for engaging randomized policies in this paper, as explained earlier,  is technical more than informational. That said, randomization will become essential also for the latter reason when we are to extend our study to involve a large investor whose actions will affect the market and hence exploration-exploitation tradeoff becomes relevant.

\subsection{An Auxiliary Problem with Gaussian Policies}
Our purpose is to develop  an approach to solve the classical Merton's problem \eqref{objective_functional_classical} subject to \eqref{eq:model factor} and \eqref{classical_wealth} by bypassing the conventional statistical estimation methods. To this end, we propose an auxiliary problem that incorporates randomized policies, and show that the solution to the original problem can be derived and computed through that of the auxiliary problem. %In particular, we will demonstrate that randomization is necessary to achieve this goal, i.e., by sampling the wealth process under stochastic policies \eqref{controlled_system}. However, we shall note that we do randomization to smoothen the objective function \eqref{objective_functional_classical} with respect to the policy, and we will revisit the exploitation-exploration tradeoff from a different perspective.

We first introduce the  following class of Gaussian (feedback) policies indexed by $\lambda \geq 0$ with a specific form of variance:
\begin{definition}
	\label{def:admissible feedback}
	A measurable, distribution-valued function $\bm\pi^{(\lambda)}: [0,T] \times \mathbb{R}_+ \times \mathbb{R} \to \mathcal{P}(\mathbb R)$, where $\lambda \geq 0$, is called an admissible policy, if
	\begin{enumerate}[(i)]
		\item $\bm\pi^{(\lambda)}(\cdot|t,w,x) = \mathcal{N}\left( \bm u(t,w,x) , \frac{\lambda}{\gamma \sigma(t,x)^2}\right)$ for some measurable function $\bm u: [0,T] \times \mathbb{R}_+ \times \mathbb{R} \to \mathbb R$, where by convention $\mathcal{N}\left( \bm u(t,w,x) , 0\right)$ is the Dirac mass at $\bm u(t,w,x)$;
		\item under $\bm\pi^{(\lambda)}$, \eqref{controlled_system} has a unique weak solution $\{W^{\bm\pi^{(\lambda)}}_t\}_{t\in [0,T]}$ satisfying $\E\left[ \sup_{0\leq t\leq T}|U(W^{\bm\pi^{(\lambda)}}_t)|  \right] < \infty$.
	\end{enumerate}
	Moreover, for a given $\lambda\geq 0$, denote the collections of all admissible policies by $\bm\Pi^{(\lambda)}$.
\end{definition}

This class of Gaussian policies is inspired by certain entropy-regularized optimization problems; see \cite{ziebart2008maximum} for a discrete-time setting and  \cite{wang2018exploration}, \cite{wang2019continuous}, and \cite{ddj2020calibrating}
for continuous-time counterparts. % Such randomized behavior can often be justified by random utility theory \citep{mcfadden2001economic} or perturbed utility representation \citep{fudenberg2015stochastic}, and many others \citep{feng2017relation}.
The variance of such a policy is inversely proportional to the volatility of the stock prices and the agent risk aversion level. The exogenous parameter $\lambda\geq0$ controls the additional randomness (arising from policy randomization) introduced to the system.

%Given an admissible policy $\bm\pi^{(\lambda)}$, \eqref{controlled_system} admits a unique solution $\{W^{\bm\pi^{(\lambda)}}_s\}_{s\in [0,T]}$. Hence, based on the special structure of the variance, we may denote as
%\begin{equation}
%\label{eq:gaussian policy}
%\bm\pi_t^{(\lambda)} =   \mathcal{N}\left( \bm u(t,W^{\bm\pi^{(\lambda)}}_t,X_t) , \frac{\lambda}{\gamma \sigma(t,X_t)^2}\right) %= \mathcal{N}\left( \tilde{\bm u}(t,W^{\bm\pi^{(\lambda)}}_t, G_t) , \lambda \gamma^{-1} G_t^{-1}\right) ,
%\end{equation}
%where $G_t$ is the observed instantaneous variance, and $\tilde{\bm u}(t,w,\sigma(t,x)^2) = \bm u(t,w,x)$. On the one hand, at the implementation stage, it suffices to know the function $\tilde{\bm u}$. On the other hand, for theoretical analysis, we keep using the notation $\bm u$ as a function of the market factor for simplicity.

%More importantly, the index $\lambda$ quantifies the concentration of the stochastic policies. Larger (smaller) $\lambda$ stands for more (less) randomness is introduced. Finally,

For any given $\lambda\geq0$, the objective of our auxiliary problem is  to maximize
\begin{equation}
	\label{eq:objective new}
	J^{(\bm\pi^{(\lambda)})}(t, w, x) =  \E \left[
	U(W^{\bm\pi^{(\lambda)}}_T)  \mid W^{\bm\pi^{(\lambda)}}_t = w, X_t = x \right] ,
\end{equation}
with the {\it optimal} value function
\begin{equation}\label{optimalvalue}
	V^{(\lambda)}(t, w, x) =  \max_{\bm \pi^{(\lambda)}\in \bm\Pi^{(\lambda)}} \E\left[
	U(W^{\bm\pi^{(\lambda)}}_T) \mid W^{\bm\pi^{(\lambda)}}_t = w, X_t = x \right] ,\;\;(t,w,x)\in[0,T] \times \mathbb{R}_+ \times \mathbb{R}.
\end{equation}
%In particular, if $\lambda$ reduces to zero, then the admissible policy degenerates to the deterministic policy in the classical case. We may define the value function associated with a policy as
Note that this auxiliary problem is different from the entropy-regularized problems studied in \cite{wang2019continuous}, where the entropy of the policy is explicitly included in the objective functional. %This suggests that we consider randomness as exogenous, rather than as part of the optimization objective. }}

\section{Theoretical Analysis}
\label{sec:theory analysis}
\subsection{Ground Truth Solution to the Auxiliary Problem}
We first answer the question on the relation between the auxiliary problem \eqref{optimalvalue} and the original one \eqref{objective_functional_classical}. It is straightforward, as in \cite{wang2018exploration}, to derive that the optimal value function $V^{(\lambda)}$ satisfies the following HJB equation via dynamic programming for \eqref{optimalvalue}:
\begin{equation}
\label{eq:hjb 0}
\begin{aligned}
& \frac{\partial V^{(\lambda)}}{\partial t} + \sup_{\bm u \in \mathbb R}\Bigg\{ \Big( r + \big(\mu(t,x) - r\big) \bm u \Big)w V^{(\lambda)}_w + \frac{1}{2}\sigma^2(t,x)\Big( \bm u^2 +  \frac{\lambda}{\gamma \sigma(t,x)^2}\Big) w^2V^{(\lambda)}_{ww} \\
& + m(t,x) V^{(\lambda)}_x + \frac{1}{2}\nu^2(t,x)V^{(\lambda)}_{xx} + \rho\nu(t,x)\sigma(t,x)\bm u wV^{(\lambda)}_{wx}  \Bigg\} = 0,
\end{aligned}
\end{equation}
with the terminal condition $V^{(\lambda)}(T,w,x) =  U(w) = \frac{w^{1-\gamma} - 1}{1-\gamma}$.

At first glance, equation \eqref{eq:hjb 0} is a highly nonlinear PDE and appears hard to analyze. However, we can reduce it to a simpler PDE based on which the optimal randomized policy can be explicitly represented.
\begin{theorem}
\label{thm:hjb solution}
Suppose $\varphi$ is a classical solution of the following PDE
\begin{equation}
\label{eq:hjb 4}
\begin{aligned}
	\frac{\partial \varphi}{\partial t} & + (1-\gamma)r  + m(t,x)\varphi_x + \frac{1}{2}\nu^2(t,x)(\varphi_{xx} + \varphi_x^2) \\
	& + \frac{1-\gamma}{2\gamma}\left[ \frac{\big( \mu(t,x) - r \big)^2}{\sigma^2(t,x)} + \frac{2\rho\big( \mu(t,x)-r \big)\nu(t,x)}{\sigma(t,x)}\varphi_x  + \rho^2\nu^2(t,x)\varphi_x^2\right]  = 0,
\end{aligned}
\end{equation}
with the terminal condition $\varphi(T,x) = 0$, and $\varphi$ satisfies the regularity condition that  $\{ e^{ (1+\epsilon) \varphi(t,X_t) } \}_{t\in [0,T]}$ is uniformly integrable for some $\epsilon > 0$. Then
\begin{equation}
\label{eq:optimal value function}
V^{(\lambda)}(t,w,x) = \frac{w^{1-\gamma} \exp\{\varphi(t,x) - \lambda (1-\gamma) (T-t)/2\} -1}{1-\gamma}
\end{equation}
is a classical solution to the HJB equation \eqref{eq:hjb 0}. Moreover,
\begin{equation}
\label{eq:optimal exploratory policy via u}
\bm\pi^{(\lambda)*}(t,x) = \mathcal{N}\left( \bm u^*(t,x), \frac{\lambda}{\gamma\sigma^2(t,x)}\right),\text{ with } \bm u^*(t,x) = \frac{\mu(t,x) - r}{\gamma \sigma^2(t,x)} +\frac{\rho\nu(t,x)}{\gamma\sigma(t,x)}\varphi_x(t,x),
\end{equation}
is the optimal policy to the auxiliary problem \eqref{optimalvalue} subject to \eqref{eq:model factor} and \eqref{controlled_system}. Furthermore, $\bm u^*$ is the optimal policy for the original Merton's problem \eqref{objective_functional_classical} subject to \eqref{eq:model factor} and \eqref{classical_wealth}.
\end{theorem}

Theorem \ref{thm:hjb solution} characterizes the optimal ground truth solution (i.e. the theoretical solution {\it assuming} all the model coefficients are known) to the auxiliary problem \eqref{optimalvalue} via the PDE \eqref{eq:hjb 4} and reveals that its mean is none other than the optimal solution of the original problem \eqref{objective_functional_classical}. Note that this is a {\it theoretical} result not to be used to compute the solutions to either problems. Rather, its importance lies in its implication: {\it one can solve \eqref{objective_functional_classical} via solving \eqref{optimalvalue}.} It in turn justifies our approach of employing  a special class of Gaussian policies to recover the optimal solution of the original Merton problem.  Moreover, Theorem \ref{thm:hjb solution} indicates that we can limit  the admissible policies to only bivariate functions $\bm u$ of $(t,x)$, thus greatly reducing the complexity in solving the auxiliary problem.

As we discussed earlier there is no informational motive to study the auxiliary problem with randomized policies due to the small investor in question (while there is such a motive in the case of a large investor whose actions impact the asset prices). Engaging the auxiliary problem \eqref{optimalvalue} is a {\it technical} approach to learn the optimal solution to the original problem, as stipulated by Theorem \ref{thm:hjb solution}.  What is more, we will show subsequently that \eqref{optimalvalue} can be solved by a policy improvement algorithm, which does {\it not} work directly on \eqref{objective_functional_classical}. %this approach is not feasible for deterministic policies, as we will demonstrate next. In particular, what we suggest is to adopt the learned mean component of the learned policy, i.e., $\bm u$, to construct the portfolio every time while updating $\bm u$ by sampling randomized decisions to solve the auxiliary problem (like in backtesting).

However, taking randomized policies is not free, because the utility value decreases due to the additional randomness borne by a risk-averse agent.
We now study this ``cost"  by  comparing them to deterministic policies (i.e. those  with $\lambda = 0$) in terms of the {\it equivalent relative wealth loss} (ERWL) defined as follows.
\begin{definition}
\label{def:erwl}
We define the equivalent relative wealth loss $\operatorname{ERWL}(\bm\pi^{(\lambda)})$ of an admissible policy $\bm\pi^{(\lambda)}$ as $\Delta=\Delta(t,x)$ satisfying
\[ J^{(\bm\pi^{(\lambda)})}(0,w,x) = V^{(0)}(0,w (1-\Delta), x) . \]
\end{definition}

So ERWL $\Delta$ is a percentage in wealth with which investor is
indifferent between obtaining the ground truth value of the optimal deterministic policy with initial endowment
$w(1-\Delta)$ and getting the value of the optimal randomized policy with initial endowment $w$. In other words, $\Delta$ is
the relative cost the investor is willing to pay to engage randomized policies.

\begin{corollary}
\label{coro:erwl}
The equivalent relative wealth loss of the $\lambda$-optimal randomized policy $\bm\pi^{(\lambda)*}$ is a constant that only depends on $\lambda$ and the length of the episode $T$. Specifically, $\operatorname{ERWL}(\bm\pi^{(\lambda)*}) = 1-\exp\{-\lambda T/2\} \approx \lambda T/2 + O(\lambda^2 T^2)$.
\end{corollary}

Corollary \ref{coro:erwl} quantifies the loss of efficiency in the (relative) monetary term due to a randomized policy: A longer investment horizon or a larger $\lambda$ incurs larger losses, which is clearly intuitive. %This result is consistent with our discussion on the conventional exploitation-exploration tradeoff in Section \ref{subsec:incentive}. Given that there are no gains in the agent's information, it seems that randomizing decisions is always worse off.

\subsection{Reinforcement Learning Methods for Solving the Auxiliary Problem}
With the class of Gaussian policies, we are in the realm of RL and thus able to apply/develop   RL methods to solve the auxiliary problem. The basic idea follows the actor-critic approach developed for general stochastic control problems in \citet{jia2021pgac}, with a major modification for Merton's problem.

An actor-critic type algorithm learns the value function and the policy function alternately and iteratively. The critic refers to the policy evaluation stage (estimating the value function under the current policy), and the actor corresponds to the policy improvement stage (updating the policy guided by the value function). Theorem \ref{thm:hjb solution} informs that it suffices to learn two bivariate functions of $(t,x)$, $\bm u^*$ for the policy and $\varphi$ for the value function. %In particular, the employed stochastic policy should adopt the particular form in Definition \ref{def:admissible feedback}, where the volatility trajectory $G_t = \sigma^2(t,X_t)$ can be observed or computed. This assumption is premised upon the well-documented results that the volatility
%may be approximated by VIX, option data, or high-frequency observation of stock returns.\footnote{For example, the instant variance by calculating the realized variance with high-frequency observations \citep{barndorff2002estimating, hansen2006realized}. Alternatively, it is also possible to use the derivative price on realized variance \citep{carr2005pricing} as a proxy for the instantaneous variance, such as VIX for S\&P 500 index.}
These bivariate functions can be approximated by, e.g., a certain parametric form, linear spans of basis functions like polynomials, or neural networks. We will specify them later in our numerical study.

Now that our task reduces to learning the two functions $\bm u^*$ and $\varphi$ that only depend on the time and the market factors, it turns out these functions possess nice properties that they are ``closed'' in the iterative procedures of policy evaluation-improvement, precisely stipulated by Theorem \ref{lemma:value function one policy}:
\begin{theorem}
\label{lemma:value function one policy}
\begin{enumerate}[(i)]
\item The value function under an admissible policy $\bm\pi^{(\lambda)}(\cdot|t,x) = \mathcal{N}\left( \bm u(t,x) , \frac{\lambda}{\gamma \sigma(t,x)^2}\right)$ can be represented as
\begin{equation} \label{Jtwx} J^{(\bm\pi^{(\lambda)})}(t, w, x) =  \frac{w^{1-\gamma} \exp\{ \bar\varphi(t,x) - \lambda (1-\gamma)(T-t)/2 \} - 1}{1-\gamma} , \end{equation}
where $\bar\varphi$ satisfies the PDE
\begin{equation} \begin{aligned}\label{PDEbar}
		\frac{\partial \bar\varphi}{\partial t} & + (1-\gamma)r  + m(t,x)\bar\varphi_x + \frac{1}{2}\nu^2(t,x)\left(\bar\varphi_{xx} + \left( \bar\varphi_x \right)^2 \right) \\
		& + (1-\gamma)\left[ (\mu(t,x) - r) \bm u(t,x) - \frac{\gamma}{2}\sigma(t,x)^2 \bm u(t,x)^2  + \rho \sigma(t,x)\nu(t,x) \bm u(t,x) \bar\varphi_x \right]  = 0,
\end{aligned} \end{equation}
with terminal condition $\bar\varphi(T,x) = 0$.
\item Define a new policy
\begin{equation}
	\label{eq:improved policy}
	\tilde{\bm\pi}^{(\lambda)}(\cdot| t,x) = \mathcal{N}\left( \tilde{\bm u}(t,x), \frac{\lambda}{\gamma \sigma(t,x)^2 } \right),\ \tilde{\bm u}(t,x) = \frac{\mu(t,x)-r}{\gamma\sigma^2(t,x)} + \frac{\rho \nu(t,x)}{\gamma \sigma(t,x)} \bar\varphi_x(t,x)  .
\end{equation}
Then this new policy $\tilde{\bm\pi}^{(\lambda)}$ improves $\bm\pi^{(\lambda)}$: %, and correspondingly, the deterministic policy $\tilde{\bm u} $ improves $\bm u$. That is,
$J^{(\tilde{\bm\pi}^{(\lambda)})}(t, w, x) \geq J^{(\bm\pi^{(\lambda)})}(t, w, x)$ for all $(t,w,x)$.
\end{enumerate}
\end{theorem}

Theorem \ref{lemma:value function one policy}-(i) confirms  that it is indeed sufficient to consider the specific form \eqref{eq:optimal value function} for the critic because {\it any} value function is of that form.  Theorem \ref{lemma:value function one policy}-(ii) entails a \textit{policy improvement} theorem, which specifies a provably better policy over any given policy. We emphasize that this theorem is a {\it theoretical} result and cannot in itself be used to compute the optimal policy and value function as we do not have access to the coefficients of the PDE \eqref{PDEbar}. %Yet it has significance implication for algorithm design because it justifies the structural forms of the critic and the actor given by \eqref{eq:form of value function general} for learning. %confirms that it is sufficient to consider the of the result  constructs a new policy that is no worse than any initial point. Importantly, our improvement applies to two different levels. The stochastic policy gets improved, and so does the correspondingly deterministic counterpart. This effect provides the justification for why we would like to learn a stochastic policy rather than a deterministic one despite the latter being desired.

The next theorem, however, forms the foundation for the algorithms we are going to devise, by characterizing the value function associated with any given admissible policy  as well as the improved policy that are both theoretically identified by Theorem \ref{lemma:value function one policy}.
\begin{theorem}
\label{thm:pg theory}
\begin{enumerate}[(i)]
\item Let {$\lambda > 0$}, an admissible Gaussian policy ${\bm\pi}^{(\lambda)}$, and a continuous function $\hat V$ be given, satisfying $\hat V(T,w,x) = U(w)$ and for every $(t_0,w_0,x_0)\in [0,T) \times \mathbb{R}_+ \times \mathbb{R}$,
\[ \E\left[ \int_{t_0}^T \xi(t,W_t^{a^{{\bm\pi}^{(\lambda)}}},X_t) \dd \hat V(t, W_t^{a^{{\bm\pi}^{(\lambda)}}}, X_t)   \right] = 0 \;\text{ for any measurable function $\xi$, } \]
where $(W^{a^{{\bm\pi}^{(\lambda)}}},X)$ is the wealth-factor process under a control $a^{{\bm\pi}^{(\lambda)}}$ sampled from ${\bm\pi}^{(\lambda)}$ with the initial $W^{a^{{\bm\pi}^{(\lambda)}}}_{t_0}=w_0$, $X_{t_0}=x_0$. Then $\hat V\equiv J^{(\bm\pi^{(\lambda)})}$, which is given in Theorem \ref{lemma:value function one policy}-(i).
\item Let $\lambda > 0$, a continuous function $\hat{\bm u}$ and its associated policy $\hat{\bm\pi}^{(\lambda)}(\cdot|t,x)  = \mathcal{N}\left(\hat{\bm u}(t,x) , \frac{\lambda}{\gamma \sigma(t,x)^2}\right) $ be given satisfying for every $(t_0,w_0,x_0)\in [0,T) \times \mathbb{R}_+ \times \mathbb{R}$,
\[ \E\left[ \int_{t_0}^T \eta(t,W_t^{a^{\hat{\bm\pi}^{(\lambda)}}},X_t) \left( a_t^{\hat{\bm\pi}^{(\lambda)}} - \hat{\bm u}(t,X_t) \right) \dd J^{(\bm\pi^{(\lambda)})}  (t, W_t^{a^{\hat{\bm\pi}^{(\lambda)}}}, X_t) \right] = 0 \;\text{ for any measurable function $\eta$, }\]
where $(W^{a^{\hat{\bm\pi}^{(\lambda)}}},X)$ is the wealth-factor process under a control $a^{\hat{\bm\pi}^{(\lambda)}}$ sampled from $\hat{\bm\pi}^{(\lambda)}$ with the initial $W^{a^{\hat{\bm\pi}^{(\lambda)}}}_{t_0}=w_0$, $X_{t_0}=x_0$. Then
$\hat{\bm u}\equiv  \tilde{\bm u}$, which is constructed in Theorem \ref{lemma:value function one policy}-(ii).
\end{enumerate}
\end{theorem}

The two equations in
Theorem \ref{thm:pg theory} are types of martingale orthogonality conditions studied extensively in \cite[Section 4.2]{jia2021policy} that lead to model-free, data-driven stochastic approximation algorithms to compute the value and policy functions by choosing appropriate classes of the ``test functions" $\xi$ and $\eta$; see the next subsection for details. %  It confirms that an improved policy can be calculated with rich test functions.
Notably, Theorem \ref{thm:pg theory}-(ii) explains why a randomized policy {\it needs} to be considered in our approach. When only deterministic policies are adopted (i.e., $\lambda=0$), the orthogonality condition in Theorem \ref{thm:pg theory}-(ii) holds trivially for any $\hat{\bm u}$ because $a_t^{\hat{\bm\pi}^{(\lambda)}}\equiv \hat{\bm u}(t,X_t)$, hence becomes useless. % Only by injecting random deviation to $a_t^{\hat{\bm\pi}^{(\lambda)}}$, can we determine whether it improves an existing policy or not.

\paragraph{{\bf Data-driven RL algorithms.}}

%Model-free RL algorithms are in general data driven, namely they learn the optimal policy based on {\it observable or computable} data. Here data include both exogenous and endogenous ones: the former are typically those of the state processes while the latter strategically generated including those by randomization. For the Merton problem considered in this paper, the state processes are $X$ and $W$.

%Given that the market environment is not affected by the agent's portfolio choices, it is natural to consider using market data and statistical methods to estimate \eqref{eq:model stock} and \eqref{eq:model factor}.

%To facilitate data-driven (possibly online) learning, we consider the algorithms that rely on the temporal difference (TD) and policy gradient. Continuous-time counterparts of these concepts have been studied in \citet{jia2021policy,jia2021pgac}.

Based on Theorem \ref{lemma:value function one policy}, we only need to learn the functions $\bar\varphi$ and $\tilde{\bm u}$ in order to determine the value function of a given randomized policy and its improved policy, respectively.
Denote by $\hat \varphi^{\psi}$ and $\hat{\bm u}^{\theta}$  the respective approximated functions of $\bar\varphi$ and $\tilde{\bm u}$, where $(\psi,\theta)$ are finite-dimensional parameters to be learned. Then the corresponding approximated  value function and  (improved) policy are
\begin{equation}
\label{eq:form of value function general}
\hat V^\psi(t,w,x) = \frac{w^{1-\gamma} \exp\{\hat \varphi^{\psi}(t,x) - \lambda (1-\gamma) (T-t)/2\} - 1}{1-\gamma},\ \hat{\bm\pi}^{\theta}(\cdot|t,x) = \mathcal{N}\left( \hat{\bm u}^{\theta}(t,x), \frac{\lambda}{\gamma \sigma^2(t,x)} \right).
\end{equation}
Applying Theorem \ref{thm:pg theory} to the above functions yields % suggest that RL methods aim to use suitable stochastic approximation algorithms to solve a system of conditions that is guided by Theorem \ref{thm:pg theory} (known as the martingale characterizations of the optimal value function and optimal policy in \citealt{jia2021policy,jia2021pgac}). More precisely,
\begin{equation}
\label{eq:rl td equation general}
\left\{
\begin{aligned}
& \E\left[\int_0^T \xi_t \dd \hat{V}^{\psi}(t,W_t^{a^{\hat{\bm\pi}^{\theta}}}, X_t)  \right] = 0, \\
& \E\left[ \int_0^T \eta_t \frac{a_t^{\hat{\bm\pi}^{\theta}} - \hat{\bm u}^{\theta}(t,X_t)}{\lambda/(\gamma G_t)}   \dd \hat{V}^{\psi}(t,W_t^{a^{\hat{\bm\pi}^{\theta}}},X_t)  \right] = 0,
\end{aligned} \right.
\end{equation}
for suitably chosen ``test processes" $\xi_t=\xi(t,W_t^{a^{\hat{\bm\pi}^{\theta}}}, X_t)$ and $\eta_t=\eta(t,W_t^{a^{\hat{\bm\pi}^{\theta}}}, X_t) $, where $a_t^{\hat{\bm\pi}^{\theta}}$ is the portfolio sampled from  $\hat{\bm\pi}^{\theta}$ at time $t$ and $W^{a^{\hat{\bm\pi}^{\theta}}}$ is the observed wealth process satisfying the wealth equation \eqref{classical_wealth} under the resulting portfolio process, and $G_t=\sigma(t,X_t)^2$. \footnote{Here, we have assumed that continuous sampling of $a_t^{\hat{\bm\pi}^{\theta}}$ is possible. In actual implementation, the integrals in \eqref{eq:rl td equation general} will be replaced by summations and the sampled wealth-factor process will be evaluated with the forward Euler scheme that requires  sampling $a_t^{\hat{\bm\pi}^{\theta}}$ only at discrete times, as illustrated in the next subsection.}

%Henceforth we assume that the volatility process $G$ is observable.
%This assumption is premised upon the well-documented results that the volatility
%may be approximated accurately by VIX, option data, or high-frequency observation of stock returns.\footnote{For example, the instant variance by calculating the realized variance with high-frequency observations \citep{barndorff2002estimating, hansen2006realized}. Alternatively, it is also possible to use the derivative price on realized variance \citep{carr2005pricing} as a proxy for the instantaneous variance, such as VIX for S\&P 500 index.}
With  specified test functions $\xi,\eta$, \eqref{eq:rl td equation general} becomes a coupled system of algebraic equations in $(\psi,\theta)$, where the coefficients can be computed by {\it observable} data. The system of equations is also known as the \textit{moment conditions} or \textit{estimating equations} in the literature of generalized method of moment \citep{hansen1982generalized} in econometrics. However, we emphasize that the critical difference between econometrics and RL lies in that  data are both exogenous and endogenous and a part of the solution with the latter, because samples of portfolios and wealth both depend on the policy $\hat{\bm\pi}^{\theta}$ that needs to be learned.

In the RL literature, typical choices of the test processes are $\xi_t = \frac{\partial }{\partial \psi}\hat V^{\psi}(t,W_t^{a^{\hat{\bm\pi}^{\theta}}}, X_t)$ and $\eta_t = \frac{\partial }{\partial \theta}\hat{\bm u}^{\theta}(t,X_t) $, leading to the so-called ``TD(0)'' type of algorithms; see  e.g., \citet{sutton2011reinforcement}. However, there is no formal theory on the ``optimal" choice of these processes. For our problem, we propose the following
\begin{equation}
\label{eq:choice of test function}
\xi_t = \frac{\frac{\partial }{\partial \psi}\hat{\varphi}^{\psi}(t, X_t)}{(1-\gamma)\hat V^{\psi}(t,W_t^{a^{\hat{\bm\pi}^{\theta}}}, X_t)  + 1},\ \eta_t = \frac{\frac{\partial }{\partial \theta}\hat{\bm u}^{\theta}(t,X_t) }{(1-\gamma)\hat V^{\psi}(t,W_t^{a^{\hat{\bm\pi}^{\theta}}}, X_t)  + 1} ,
\end{equation}
which effectively replace the TD error term $\dd \hat V^{\psi}_t$ with an adjusted, ``relative'' TD error $\frac{\dd \hat V^{\psi}_t}{(1-\gamma) \hat V^{\psi}_t + 1}$ in a conventional TD(0) algorithm.
The reason for this adjustment  is due to the homothetic property of the CRRA utility function. In particular, the wealth processes are typically growing and non-stationary, which may cause instability in the learning process. %On the other hand, the optimal policy is wealth-independent, which makes the wealth itself a redundant variable.
The purpose of the denominator in \eqref{eq:choice of test function} is to normalize the wealth effect. %Note that such elimination is only on the functional form of the learning signals, and it does not mean we ignore the market variables. All the useful information will be carried along through the stock returns, as we will show next with the Black-Scholes model.

%Theorem \ref{thm:pg theory} connects the orthogonality conditions \eqref{eq:rl td equation general} to the policy improvement theorem in Lemma \ref{lemma:value function one policy} $(ii)$ under the assumption of an oracle that can calculate the value function $J^{(\bm\pi^{(\lambda)})}$ under any admissible policy $\bm\pi^{(\lambda)}$ accurately\footnote{In the typical paradigm of actor-critic learning, one shall first conduct a policy evaluation to obtain a relatively accurate estimate of the value function and then perform the policy gradient to find an improved policy. Hence, this assumption is aligned with the typical practice of actor-critic learning.}. It confirms that the improved policy can be calculated with rich test functions.

\subsection{The Impact of Temperature Parameter: The Worked-Out Example of BS Market}\label{BS}

To illustrate the general results derived so far and to further investigate the impact of the temperature parameter $\lambda$, we consider the classical Black-Scholes market where there are a risk-free asset and a risky one with constant model coefficients, and  there is no market factor. Theorem \ref{thm:hjb solution} then specializes  to a simple solution with $\bm u^* = \frac{\mu - r}{\gamma \sigma^2}$ and $V^{(\lambda)}(t,w) = \frac{w^{1-\gamma}\exp\{ [r + \frac
{(\mu-r)^2}{2\gamma\sigma^2}](1-\gamma)(T-t) -\lambda(1-\gamma)(T-t)/2 \} - 1}{1-\gamma}$. Once again, agent's knowledge includes the values of $\gamma,T$ and $\sigma$, whereas $\lambda$ is a fixed temperature parameter.  In particular, the mean return $\mu$ is unknown. This setting is consistent with a consensus that the stock  expected return is more difficult if not impossible to estimate accurately using  statistical methods; see e.g. \cite{luenbergerinvestment}.

Inspired by the (theoretical) ground truth optimal solution, we
approximate the value and policy functions with two scalar parameters $\psi$ and $\theta$:
\[ \hat V^{\psi}(t,w) = \frac{w^{1-\gamma} \exp\{ \psi(T-t) - \lambda (1-\gamma)(T-t)/2\} - 1}{1-\gamma},\ \hat{\bm u}^{\theta}(t) = \theta,\ \hat{\bm \pi}^{\theta}(\cdot|t) = \mathcal{N}\left(\theta, \frac{\lambda}{\gamma\sigma^2}\right) . \]

With the proposed test processes \eqref{eq:choice of test function}, the optimality conditions \eqref{eq:rl td equation general} now read
\begin{equation}
\label{eq:black scholes foc}
\left\{ \begin{aligned}
& \E\left[  \int_0^T \frac{T-t}{(W_t^{a^{\hat{\bm\pi}^{\theta}}})^{1-\gamma} \exp\{\psi (T-t) - \lambda (1-\gamma)(T-t)/2\}} \dd \frac{(W_t^{a^{\hat{\bm\pi}^{\theta}}})^{1-\gamma}\exp\{\psi (T-t) - \lambda (1-\gamma)(T-t)/2\} }{1-\gamma}  \right] = 0, \\
& \E\left[  \int_0^T \frac{\gamma \sigma^2( a_t^{\hat{\bm\pi}^{\theta}}- \theta)}{\lambda(W_t^{a^{\hat{\bm\pi}^{\theta}}})^{1-\gamma} \exp\{\psi (T-t) - \lambda (1-\gamma)(T-t)/2\}} \dd \frac{(W_t^{a^{\hat{\bm\pi}^{\theta}}})^{1-\gamma}\exp\{\psi (T-t) - \lambda (1-\gamma)(T-t)/2\}}{1-\gamma}  \right] = 0.
\end{aligned}
\right. 
\end{equation}

\paragraph{{\bf Informal analysis: optimality conditions.}}
To better understand the conditions \eqref{eq:black scholes foc}, we first present an informal analysis by ignoring the time discretization issue (i.e.  assuming it is possible to
continuously draw samples from a randomized policy and collect observations, and to compute the integrals involved  exactly).

Applying It\^o's lemma to the term $\dd \frac{(W_t^{a^{\hat{\bm\pi}^{\theta}}})^{1-\gamma}\exp\{\psi (T-t) - \lambda (1-\gamma)(T-t)/2\}}{1-\gamma} $  and using the wealth equation \eqref{classical_wealth}, we deduce  that \eqref{eq:black scholes foc} is equivalent to
{\small \begin{equation}
\label{eq:black scholes foc simplify}
\left\{
\begin{aligned}
	& \E\left[ \int_0^T (T-t) \left( -\frac{\psi}{1-\gamma} + \frac{\lambda}{2} + r + (\mu-r)a_t^{\hat{\bm\pi}^{\theta}} - \frac{\gamma}{2}\sigma^2 (a_t^{\hat{\bm\pi}^{\theta}})^2 \right) \dd t    + \int_0^T(T-t) \sigma a_t^{\hat{\bm\pi}^{\theta}} \dd B_t \right] = 0,\\
	& \E\left[ \int_0^T \frac{\gamma\sigma^2}{\lambda}(a_t^{\hat{\bm\pi}^{\theta}} - \theta) \left( -\frac{\psi}{1-\gamma} + \frac{\lambda}{2} + r + (\mu-r)a_t^{\hat{\bm\pi}^{\theta}} - \frac{\gamma}{2}\sigma^2 (a_t^{\hat{\bm\pi}^{\theta}})^2 \right) \dd t    + \int_0^T \frac{\gamma\sigma^2}{\lambda}(a_t^{\hat{\bm\pi}^{\theta}} - \theta)  \sigma a_t^{\hat{\bm\pi}^{\theta}} \dd B_t \right] = 0.
\end{aligned}
\right. 
\end{equation}}

Because  $a_t^{\hat{\bm\pi}^{\theta}} \sim \mathcal{N}(\theta, \frac{\lambda}{\gamma\sigma^2})$, the expectations in \eqref{eq:black scholes foc simplify} can be explicitly calculated, yielding the following system of equations:
\begin{equation}
\label{eq:black scholes foc simplify final}
\frac{T^2}{2}\left( -\frac{\psi}{1-\gamma} + r + (\mu-r)\theta - \frac{\gamma}{2}\sigma^2\theta^2 \right)  = 0,\  \ T \gamma\sigma^2\left( \frac{\mu-r}{\gamma\sigma^2} - \theta\right) = 0.
\end{equation}
The solutions to these equations coincide with the theoretical ground truth, which in turn implies that the optimality conditions  \eqref{eq:black scholes foc simplify} or \eqref{eq:black scholes foc} indeed lead to the {\it correct} solutions.
Meanwhile, in this special case, \eqref{eq:black scholes foc simplify final} shows that the second equation regarding policy optimization (in $\theta$) is decoupled from the first equation on policy evaluation (in $\psi$ and $\theta$). Since our primary interest lies in finding the optimal policy, we shall therefore focus on the second equation only.

If the second equation of \eqref{eq:black scholes foc simplify} can be computed and implemented perfectly without the need of discretization, it will yield the correct optimal policy solution, $\theta$, immediately.
However, the above  informal analysis raises a puzzling question: What is the impact of the randomization measured by the temperature parameter $\lambda$? The second equation in \eqref{eq:black scholes foc simplify final} appears to be independent of $\lambda$. So, why do we still need to randomize with $\lambda>0$? To resolve this puzzle, we must conduct a formal analysis from the sampling perspective.

\paragraph{{\bf A formal analysis: impacts of discretization and randomization.}}
We now present a formal analysis by incorporating sampling errors in our procedure.

For numerical implementation, the term inside the expectation, say $e(\psi,\theta)$, in the second equation of \eqref{eq:black scholes foc} needs to be estimated by samples collected with suitable time discretization for the integration and discrete sampling of the randomized policy. % $a_t^{\hat{\bm\pi}^{\theta_n}}, W_t^{a^{\hat{\bm\pi}^{\theta_n}}}$.
More precisely, for equally spaced time grids $0=t_0 < t_1<\cdots < t_K = T$ with grid size $\Delta t = \frac{T}{K}$ and a given set of value and policy function parameters $(\psi,\theta)$, we denote  by $\widehat{e(\psi,\theta)}$ the estimate of $e(\psi,\theta)$, computed by
\begin{equation}
\label{eq:increment samples discrete}
\begin{aligned}
\widehat{e(\psi,\theta)} = \sum_{k=0}^{K-1} & \frac{\gamma \sigma^2 (a_{t_k} - \theta)}{\lambda (W_{t_k})^{1-\gamma} \exp\{ \psi(T-t_k) - \lambda(1-\gamma)(T-t_k)/2  \}} \frac{1}{1-\gamma} \\
& \times \bigg[ (W_{t_{k+1}})^{1-\gamma} \exp\{ \psi(T-t_{k+1}) - \lambda(1-\gamma)(T-t_{k+1})/2  \}  \\
& - (W_{t_k})^{1-\gamma} \exp\{ \psi(T-t_k) - \lambda(1-\gamma)(T-t_k)/2  \}\bigg] \\
= \sum_{k=0}^{K-1} & \frac{\gamma \sigma^2 (a_{t_k} - \theta)}{\lambda (1-\gamma)} \left[ \left( \frac{W_{t_{k+1}}}{W_{t_{k}}} \right)^{1-\gamma} \exp\left\{ \left[-\psi + \lambda(1-\gamma)/2 \right]\Delta t \right\}     - 1\right] ,
\end{aligned}
\end{equation}
where $a_{t_k}|W_{t_k} \sim \mathcal N(\theta, \frac{\lambda}{\gamma \sigma^2})$, and on $ (t_k,t_{k+1})$, $W$ satisfies the wealth equation \eqref{classical_wealth} with a constant portfolio  $a_{t_k}$, i.e.,
\[ \frac{\dd W_t}{W_t} = a_{t_k} \frac{\dd S_t}{S_t} + (1-a_{t_k})r\dd t. \]

The impact of the time discretization is characterized by the following proposition.
\begin{proposition}
\label{prop:gbm discrete time}
Suppose $\lambda>0$ and $\Delta t < \min\{T, \frac{1}{\theta^2}, \frac{1}{|\psi|} \frac{1}{4 \lambda |\gamma - 1|} \}$. Then there exists a constant $C$ that depends only on $\mu,r,\sigma,\gamma,T$ such that
\[ \left| \E\left[  \widehat{e(\psi,\theta)}  \right]  - T\gamma\sigma^2(\theta^*-\theta) \right| \leq  C (1 + |\theta^2| +|\psi| + \lambda) \Delta t ,\]
where $\theta^* = \frac{\mu-r}{\gamma\sigma^2}$. Moreover,
\[
\operatorname{Var}\left[\widehat{e(\psi,\theta)} \right] \leq C\left(1 + \frac{\theta^2}{\lambda}\right) + C\left(\frac{1+\psi^2+\theta^4}{\lambda}+\lambda\right) \Delta t .\]
\end{proposition}

While Proposition \ref{prop:gbm discrete time} confirms our derivation of the theoretical equivalence between the second equations of \eqref{eq:black scholes foc} (or \eqref{eq:black scholes foc simplify}) and \eqref{eq:black scholes foc simplify final}, it also shows there is a bias when we numerically compute the former due to time discretization  and it gives an upper bound of  the bias in terms of the grid size $\Delta t$ and  the strength of the randomization $\lambda$ along with other parameters.
The bound is linear in $\Delta t$, consistent with the typical rate in numerical methods for simulating SDEs and computing integrals. It is also linear in $\lambda$; so a smaller level of randomization  reduces the bias. % even though the mean of the policy is unchanged. %The reason is that more deviate portfolio choices in either directions increases the volatility of the portfolio returns, and the estimating equation is based on a risk-averse utility function.
Thus, from the bias-reducing perspective, besides a finer grid size, a smaller level of policy randomization helps.

On the other hand, Proposition \ref{prop:gbm discrete time} gives an upper bound for the variance of the learning signal samples. This bound will not vanish even if $\Delta t$ shrinks to 0, implying  that it is impossible to accurately compute the desired  quantity  with just a few  trajectories (i.e. with a small dataset) even when continuous sampling is possible. The leading term consists of  a constant and a term of the order $\lambda^{-1}$; so when the dataset size is small an elevated  level of policy randomization helps reduce the variance. %In the continuous sampling case, a purely randomized  ($\lambda=\infty$) policy yields the most accurate learning signals. However, given that discretization is inevitable, a larger $\lambda$ also increase the overall randomness of the samples but this effect only has scale $\Delta t$.

The above analysis shows that randomization is indeed relevant for learning the optimal policy parameter $\theta$ in actual implementation. Despite the results in Proposition \ref{prop:gbm discrete time} are all provided in the form of upper bounds, these upper bounds already inform us important bias-variance tradeoff of the learning signals from two aspects. First, the time discretization $\Delta t$ should shrink to zero to ensure an unbiased estimate of the learning signal, whereas diminishing time discretization cannot offset the intrinsic randomness in the signal. Second, a lower temperature $\lambda$ induces smaller bias but enlarges the variance. Therefore, we should not pick a too large or too small $\lambda$ in order to balance bias and variance. For a fixed $\lambda>0$, the next theorem gives an algorithm to compute $\theta$ based on the discretized version of the second equation of \eqref{eq:black scholes foc} along with its convergence rate and the error bound of the expected equivalent relative wealth loss.

\begin{theorem}
\label{thm:bs converge}
Fix $\lambda>0$ and consider the following update rule for the policy parameter  $\theta$:
\begin{equation}\label{poit}
\theta_{n+1} = \Pi_{[-c_{n+1}, c_{n+1}]}\left( \theta_{n} + \ell_n\widehat{e_n(\psi,\theta)}  \right), \end{equation}
where $\widehat{e_n(\psi,\theta)}$ is a sample given by \eqref{eq:increment samples discrete} with grid size $\Delta t_n$ and actor-critic parameters $(\theta_n,\psi_n)$, and $\Pi_K(\cdot)$ is the projection mapping onto a closed, convex set $K$. Assume that for any given $\eta_1 > 0, \eta_2 \in (0,1)$, there exist $n_0\in \mathbb N$ and $M>0$ such that for all $n\geq n_0$, $\ell_n = \frac{(1+\eta_1)}{(n + \eta_1)\eta_2\eta_1} < \frac{1}{T\gamma\sigma^2}$, $\Delta t_n \leq T \ell_n$, $c_{n} = \sqrt{\log n}$, and $|\psi_n| \leq M$. Then there exist constants $C_1,C_2,C_3$ that depend only on $\mu,r,\sigma,\gamma,T,n_0,M$, such that
\[ \E\left[(\theta_{n+1}-\theta^*)^2\right] \leq C_1 \ell_n \log n\sim O\left( n^{-1}\log n \right), \]
and
\[ \E\left[ \operatorname{ERWL}(\hat{\bm u}^{\theta_n}) \right] \leq \left( \max\{ \frac{C_3}{1-\eta_2} \sup_{n\geq n_0} \frac{1 + \lambda^{-1}\log n + \lambda}{\log (n-1)}   , C_2\} \right) \ell_n \log n \sim O\left( n^{-1}\log n \right) . \]
\end{theorem}

\begin{figure}
\centering
\includegraphics[width=0.5\linewidth]{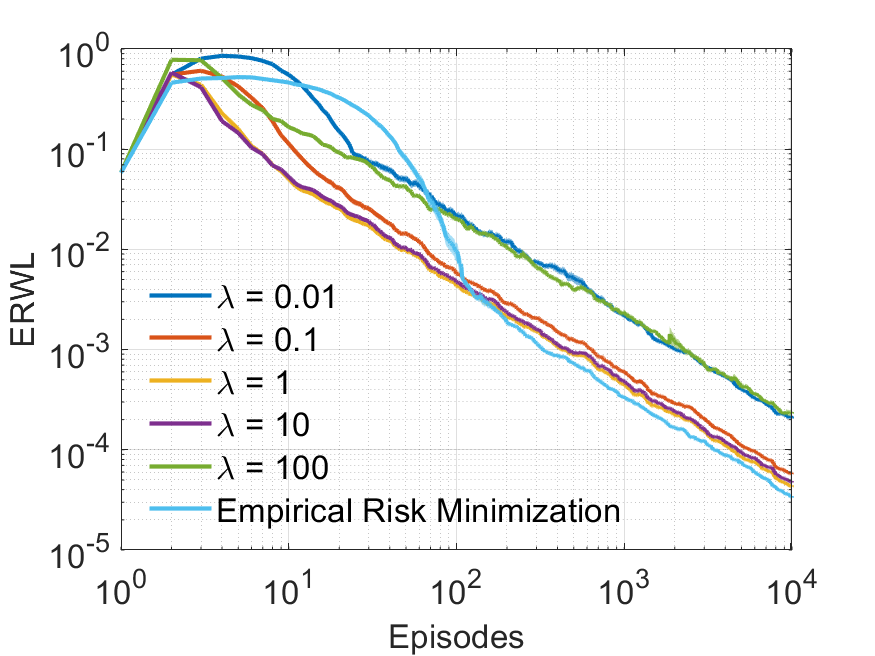}
\caption{\textbf{The comparison between the empirical risk minimization and the proposed method.} The horizontal (the number of episodes) and vertical (expected relative wealth loss) axes are both in log-scale. The shaded areas indicate the standard deviations of the estimated ERWLs. The results are based on 1000 times of independent simulation runs and 10,000 episodes of 1-year trajectory is used in each run. The model parameters are $\mu = 0.2,r=0.02,\sigma=0.3,\gamma=3,T=1$. The learning rate is $a_n = 10/(n+1)$ and the initial policy parameter is $\theta_0 = 0$. The projected region is taken as $c_n = \max\{10, \sqrt{\log (n+1)} \}$ and discretization size is $\Delta t_n = \min\{ 0.001, 10/(n+1) \}$.}
\label{fig:gbm lambda}
\end{figure}

So, under the policy iteration algorithm \eqref{poit} along with the chosen hyperparamters specified in the theorem, the algorithm converges and the expected equivalent relative wealth loss of the resulting RL policy converges to 0. Moreover, the $L^2$-convergence rate of the former and the convergence rate of the latter are both $O\left( n^{-1}\log n \right)$, matching the typical optimal convergence rate (ignoring the logarithmic factor). This asymptotic rate holds for any fixed temperature $\lambda>0$. In particular, we obtain an error bound of the expected loss in terms of the strength of the randomization, which is of the order $\lambda^{-1}$. This result reconciles with Proposition \ref{prop:gbm discrete time}  in terms of the variance of the learning signal. We illustrate the convergence rate under different values of $\lambda$ in Figure \ref{fig:gbm lambda}. In the log-log scale plot, all curves eventually exhibit the similar slope around $-1$, confirming the theoretical rate of convergence and the fact that error is reduced with a larger $\lambda$ with small sample size. Moreover, it also confirms the implications in Proposition \ref{prop:gbm discrete time} that the impact of temperature is not monotone as a high temperature would also reduce the performance.

Finally,  note that our method does not permit $\lambda = 0$ and therefore always learns a randomized policy. Alternative data-driven approaches, such as empirical risk minimization (ERM), have been developed to directly learn a deterministic policy. We provide the formulation of ERM to the Merton problem in Appendix \ref{appendix:erm} and implement it to compare with our method under the same settings. Figure \ref{fig:gbm lambda} shows that, with a small size dataset (i.e. fewer than 100 episodes), ERM performs poorly compared with our method using $\lambda=0.01,0.1,1,10,100$. Only when the dataset size is large does its performance become comparable to our method with $\lambda=0.1,1,10$, and the convergence rate is similar to ours.

%To compare  ERM with ours, we consider the Black--Scholes market with the same setting as in Section \ref{BS} where there is no market factor. In this case, $\bm u^{\theta}$ degenerates into a scalar $\theta$. While updating $\theta$, we apply the same projection and learning rates in our proposed methods in Section \ref{BS}. Figure \ref{fig:gbm erm} shows the result in terms of ERWL. It is seen that with a small size dataset (i.e. fewer than 100 episodes) ERM performs poorly compared with our methods using $\lambda=0.01,0.1,1$. Only when the dataset size is large is its performance comparable to our method with $\lambda=1$ and the convergence rate is similar to ours.

\section{A Market with Stochastic Volatility}
\label{sec:sv}
In this section, we present a stochastic volatility market environment, which is the setting for our subsequent numerical experiments, and discuss the advantages of RL over the classical plug-in approach.

A  stochastic volatility model is specified by $\mu(t,x) = r + \delta x^{\frac{1+\alpha}{2\alpha}}$, $\sigma(t,x) = x^{\frac{1}{2\alpha}}$, $m(t,x) = \iota(\bar{x} - x)$, and $\nu(t,x) = \bar{\nu}\sqrt{x}$, where $\alpha \neq 0$. This is a fairly general model studied in \cite{liu2007portfolio} for the classical utility maximization problem and  in \cite{dai2018dynamic} for the classical equilibrium mean-variance problem for log returns.
As before, we assume volatility process is observable/estimatable  and $\gamma$ is known, while
the agent has no access to any other parameter values.

\subsection{Classical Benchmark}
For readers' convenience, we review the relevant results of the classical model-based benchmark with $\lambda = 0$. The following lemma is taken from \cite{liu2007portfolio}. %Note that \eqref{eq:classical solution ricatti} suggests one way  to parameterize the value function and policy, which will be used in our numerical experiments.
\begin{lemma}
\label{lemma:classical solution}
Assuming the model primitives of the classical benchmark model satisfy
\begin{equation}\label{parameter inequality}
\iota^2\gamma > (1-\gamma)(2\rho\iota\delta\bar{\nu} + \delta^2\bar{\nu}^2),
\end{equation}
the optimal strategy is
\[ \bm u^*(t,x) =  \left(\frac{\delta}{\gamma} + \frac{\rho\bar{\nu}}{\gamma}A_1(t) \right) x^{\frac{\alpha-1}{2\alpha}} = \left(\frac{\delta}{\gamma} + \frac{\rho\bar{\nu}}{\gamma}A_1(t) \right) (\sigma^2(t,x))^{\frac{\alpha - 1}{2}}, \]
and the optimal value function is
$$V^{(0)}(t,w,x) = \frac{w^{1-\gamma}e^{A_1(t)x + A_0(t)} - 1}{1-\gamma},$$ where $A_1,A_0$ respectively satisfy the following ordinary differential equations (ODEs):
\begin{equation}
\label{eq:classical solution ode general}
\begin{aligned}
	& A_1' - \iota A_1 + \frac{1}{2}\bar{\nu}^2A_1^2 + \frac{1-\gamma}{2\gamma}\big[ \delta^2 + 2\rho\delta\bar{\nu}A_1 + \rho^2\bar{\nu}^2A^2_1 \big] = 0,\ A_1(T) = 0,\\
	& A'_0 + (1-\gamma)r - \beta + \iota \bar{x}A_1 = 0,\ A_0(T) = 0.
\end{aligned}
\end{equation}
\end{lemma}

Indeed, the two ODEs in \eqref{eq:classical solution ode general}, under the condition \eqref{parameter inequality},  can be explicitly solved with the following solutions:
\begin{equation}
\label{eq:classical solution ricatti}
\begin{aligned}
& A_1(t) = \frac{-\psi_1 + \psi_1 e^{\psi_0 (T-t)}}{-\psi_2 + \psi_3 e^{\psi_0(T-t)}},\\
& A_0(t) = \psi_4 (T-t) + \psi_5\log\frac{\big(-\psi_2+\psi_3e^{\psi_0(T-t)}\big)}{-\psi_2 + \psi_3},
\end{aligned}
\end{equation}
where
\[ \begin{aligned}
\psi_0 = & -\frac{\sqrt{\iota^2 \gamma - (1-\gamma) \delta\bar\nu(\delta\nu + 2\iota\rho)}}{\sqrt{\gamma}}, \ \ \
\psi_1 = \frac{(1-\gamma)\delta^2}{\bar{\nu}^2[\rho^2 + \gamma(1-\rho^2)]} ,\\
\psi_{2,3} = & \frac{\iota \gamma - (1-\gamma)\delta\bar\nu\rho \pm \sqrt{\gamma} \sqrt{\iota^2 \gamma - (1-\gamma) \delta\bar\nu(\delta\nu + 2\iota\rho)} }{\bar{\nu}^2[\rho^2 + \gamma(1-\rho^2)]} ,\\
\psi_4 = &(1-\gamma)r  + \iota\bar x\psi_3,\ \psi_5 = \frac{2\gamma\iota\bar x}{\bar{\nu}^2[\rho^2 + \gamma(1-\rho^2)]}  .
\end{aligned}
\]
%In \eqref{eq:explode ode},
%\[ \psi_0 = \frac{\iota \gamma - (1-\gamma)\delta\bar\nu\rho }{\bar{\nu}^2[\rho^2 + \gamma(1-\rho^2)]},\ \psi_1 = \frac{\sqrt{\gamma}\sqrt{-\iota^2 \gamma + (1-\gamma) \delta\bar\nu(\delta\nu + 2\iota\rho)}}{\bar{\nu}^2[\rho^2 + \gamma(1-\rho^2)]} ,\ \psi_2 = \frac{\sqrt{-\iota^2 \gamma + (1-\gamma) \delta\bar\nu(\delta\nu + 2\iota\rho)}}{2\sqrt{\gamma}},\]
%\[\psi_4 = (1-\gamma)r - \beta -\iota \bar{x} \psi_0,\ \psi_5 = -\frac{2\gamma\iota\bar{x}}{\bar{\nu}^2[\rho^2 + \gamma(1-\rho^2)]}.  \]

The above analytical representations require specifications of the model parameters and, hence, cannot be used directly in our RL setting. However, they give specific functional {\it structures} of the value function and policies that are helpful for function approximations. We will employ them in our subsequent numerical experiments.

%A similar result is also obtained in \cite{liu2001dynamic}, where we could find the closed-form solution to the classical asset allocation problem in a general stochastic volatility environment, if model primitives are provided.
Lemma \ref{lemma:classical solution} shows that the optimal policy can be represented as a function of the stock volatility. Moreover,  the elasticity of the instantaneous variance on the optimal policy is a constant,  given by
\[ \frac{\partial \bm u^*}{\partial \sigma^2}(t,x)\Big/\frac{\bm u^*(t,x)}{\sigma^2(t,x)} = \frac{\alpha - 1}{2}, \]
which represents the sensitivity of the portfolio with respect to the current (observed) volatility. %\dyc{The second feature of the optimal policy is that the hedging demand is reflected through the sensitivity of the value function with respect to the stochastic factor. Namely, value function provides inherent guidance on the optimal policy.}
%Note that we have a technical assumption \eqref{parameter inequality} in Lemma \ref{lemma:classical solution} to ensure the solvability of the Riccati equation, which also appears in \cite{kraft2005optimal}. This assumption implies that the risk aversion parameter should be compatible with the model primitives. This makes the classical problem fragile in some sense and suggests the advantage of the data-driven RL method, see Section \ref{sec:fragility}.

\subsection{Pitfall of Model-Based Solution and Virtue of Reinforcement Learning}\label{sec:fragility}
It is well known that solutions to  Merton's problems are highly sensitive to model primitives \citep{merton1980estimating},  especially for the stochastic volatility model. The pitfall of the traditional model-based, first-estimate-then-optimize paradigm is twofold. First, the optimal solution depends on model primitives in a highly nonlinear way, as exemplified  by \eqref{eq:classical solution ricatti}, where $\psi_i$ are complicated functions of the model parameters. This calls for an extremely accurate estimation of these functions, which  may require unrealistically long historical data. %; see our numerical experiment for an example.
%Therefore, such a plug-in policy may yield significant error and bias even if model primitive can be estimated accurately (which is not likely given realistic amount of data).}
%This type of error is well known in statistics.
Second, there is a technical assumption \eqref{parameter inequality} in Lemma \ref{lemma:classical solution}, which also appears in \cite{kraft2005optimal}. %This assumption implies that the risk aversion parameter should be compatible with the model primitives.
%in the classical case, it is worthwhile noticing that we have restriction \eqref{parameter inequality} on model primitives.
Such an assumption is to theoretically ensure the ODE system \eqref{eq:classical solution ode general} to be well-posed. When the assumption is violated, the solutions of \eqref{eq:classical solution ode general}  have completely different forms:\footnote{{Because the ODE \eqref{eq:classical solution ode general} is autonomous and separable, its general solution can be written as an indefinite integral: $T - t = \int_{A(t)}^0   \frac{\dd z}{ \iota z - \frac{1}{2}\bar{\nu}^2 z^2 - \frac{1-\gamma}{2\gamma}[\delta^2 + 2\rho\delta\bar\nu z +\rho^2\bar{\nu}^2z^2]}$. The form of the solution depends drastically on whether or not the quadratic algebraic equation $\iota z - \frac{1}{2}\bar{\nu}^2 z^2 - \frac{1-\gamma}{2\gamma}[\delta^2 + 2\rho\delta\bar\nu z +\rho^2\bar{\nu}^2z^2] = 0$ has real roots.}}
\begin{equation}
\label{eq:explode ode}
\begin{aligned}
& A_1(t) = -\psi_0 + \psi_1 \tan\big(\psi_2 (T-t) + \arctan\frac{\psi_0}{\psi_1} \big),\\
& A_0(t) = \psi_3 (T-t) + \psi_4\log\frac{\cos\big(\arctan\frac{\psi_0}{\psi_1}+\psi_2(T-t)\big)}{\cos\left( \arctan\frac{\psi_0}{\psi_1}\right)},
\end{aligned}
\end{equation}
where
\[ \begin{aligned}
\psi_0 = & \frac{\iota \gamma - (1-\gamma)\delta\bar\nu\rho }{\bar{\nu}^2[\rho^2 + \gamma(1-\rho^2)]},\ \ \
\psi_1 = \frac{\sqrt{\gamma}\sqrt{-\iota^2 \gamma + (1-\gamma) \delta\bar\nu(\delta\nu + 2\iota\rho)}}{\bar{\nu}^2[\rho^2 + \gamma(1-\rho^2)]} ,\\ \psi_2 = &\frac{\sqrt{-\iota^2 \gamma + (1-\gamma) \delta\bar\nu(\delta\nu + 2\iota\rho)}}{2\sqrt{\gamma}},\ \ \
\psi_3 =  (1-\gamma)r - \beta -\iota \bar{x} \psi_0,\ \ \
\psi_4 = -\frac{2\gamma\iota\bar{x}}{\bar{\nu}^2[\rho^2 + \gamma(1-\rho^2)]}. \end{aligned} \]
These functions will blow up to infinity periodically and thus do not lead to reasonable investment strategies. Yet, even if the true underlying market processes do satisfy \eqref{parameter inequality},  standard estimation procedures do  not usually account for such a nonlinear and nonconvex constraint. As a consequence, the {\it estimated} model primitives may violate \eqref{parameter inequality} so that the corresponding ODE system \eqref{eq:classical solution ode general} may have solutions not in the same form as \eqref{eq:classical solution ricatti}, and the resulting investment strategies may generate infinite leverage yielding  infeasible numerical computations.

By contrast, RL bypasses model estimation and learns the optimal policy {\it directly}, thereby avoiding the blow-up solutions described above that arise from the traditional plug-in method. What RL learns or estimates is now the optimal policy itself rather than model primitives,  based on performance rather than statistical properties. Specifically for the current stochastic volatility model, RL first determines the structures of the optimal policy and the value function through theoretical analysis, and then learns/updates the parameters in \eqref{eq:classical solution ricatti} through data and standard RL procedures such as policy evaluation and policy improvement.

We emphasize  the importance of exploiting the special structure of a given problem for RL algorithm design. For instance, for the present problem, it follows from both the general result (Theorem \ref{thm:hjb solution}) and the special one (Lemma \ref{lemma:classical solution}) that we only need to consider the following  class of Gaussian policies:
\begin{equation}
\label{eq: explore optimal solution asymptotic lambda general}
\operatorname{Var}(\bm\pi^*(t,x)) = \frac{\lambda }{\gamma \sigma^2(t,x)},\ \operatorname{Mean}(\bm\pi^*(t,x)) = \left[ \frac{\delta}{\gamma}+\frac{\rho \bar{\nu} }{\gamma} A_1(t)\right] (\sigma^2(t,x))^{\frac{\alpha-1}{2}},
\end{equation}
which we intentionally express in terms of the instantaneous variance  $\sigma^2$. Note that the policy variance depends only on $\sigma^2$, whereas the mean depends on $\sigma^2$ as well as other model primitives through $A$, a function of time $t$ only.
Due to this special structure obtained through theoretical analysis, we can determine the policy variance without incurring any extra training or estimation so long as a proxy of $\sigma^2$ is available/observable.  Naturally, we still need to learn  the mean of the policy, but the learning will be amply  simplified.

\subsection{Numerical Procedure}
\label{sec:numerical methods}

By It\^o's formula, the instantaneous variance  process $G_t: = \sigma^2(t,X_t) = X_t^{1/\alpha}$ satisfies
\[ \dd G_t = \bigg[\big( \frac{\iota \bar{x}}{\alpha} + \frac{(1-\alpha)\bar{\nu}^2}{2\alpha^2}\big)  G_t^{1-\alpha} - \frac{\iota}{\alpha}G_t \bigg]\dd t + \frac{\bar{\nu}}{\alpha} G_t^{1 - \alpha/2} [ \rho \dd B_t + \sqrt{1-\rho^2}\dd \tilde{B}_t]. \]
We now replace $X$ with $G$ as a state variable  (the other state variable is wealth $W$) which is  observable. %This assumption is premised upon the well-documented results that the volatility  may be  approximated by VIX, option data, or high-frequency observation of stock returns with high accuracy.
In the current setting we do not need to assume the factor $X$ to be observable.\footnote{Note that knowing the instantaneous variance $G_t$ is not equivalent to knowing the market factor $X_t$ because $\alpha$ is unknown.}
On the one hand, to our best knowledge, we do not know of any statistical method tailored for estimating the coefficients of \eqref{eq:model stock} and \eqref{eq:model factor} using time series  $\{(S_t,G_t)\}_{t\in [0,T)}$, except for the na\"ive MLE method that demands huge computational cost and large amount of data to reach desired accuracy. By contrast, RL methods take $G_t$ as inputs to learn directly the function approximators for optimal policy without having to estimate the market model.

Consequently, in view of \eqref{eq:form of value function general}, we now consider the approximated value function and policy as
\begin{equation*}
%\label{eq:form of value function general}
\hat V^\psi(t,w,g) = \frac{w^{1-\gamma} \exp\{\hat \varphi^{\psi}(t,g) - \lambda (1-\gamma) (T-t)/2\} - 1}{1-\gamma},\ \hat{\bm\pi}^{\theta}(\cdot|t,g) = \mathcal{N}\left( \hat{\bm u}^{\theta}(t,g), \frac{\lambda}{\gamma g} \right),
\end{equation*}
where the argument $g$ stands for the observable instantaneous variance $G_t = \sigma(t,X_t)^2$.

There are two ways to further parameterize these actor-critic  functions.
Inspired by Lemma \ref{lemma:classical solution}, especially the expressions in \eqref{eq:classical solution ricatti}, we can parameterize the value function of a given policy as
\[ \hat V^{\psi}(t,w,g) = \frac{w^{1-\gamma}}{1-\gamma}\exp\left( A_1^{\psi}(t) g^{\psi_6} + A_0^{\psi}(t) - \frac{\lambda(1-\gamma)(T-t)}{2}  \right)  - \frac{1}{1-\gamma}, \]
where
\begin{equation}
\label{eq:parametrize value function}
A_1^{\psi}(t) = \frac{-\psi_1 + \psi_1 e^{\psi_0 (T-t)}}{\psi_2 + \psi_3 e^{\psi_0(T-t)}},\ A_0^{\psi}(t) =  \psi_4 (T-t) + \psi_5\log\frac{\psi_2+\psi_3e^{\psi_0(T-t)}}{\psi_2 + \psi_3},
\end{equation}
with $\psi\in \mathbb{R}^7$ whose components are $\psi_0,\psi_1,\cdots,\psi_6$. %This functional form is motivated by the solution to the Ricatti equation in the classical problem given by \eqref{eq:classical solution ricatti}.
Moreover, in view of both Theorem \ref{thm:hjb solution} and Lemma \ref{lemma:classical solution}, we parameterize the policy by
\[ \hat{\bm\pi}^{\theta}(a|t,g) = \frac{1}{\sqrt{ \frac{2\pi\lambda}{\gamma g}}}\exp\left\{ -\frac{\gamma g}{2 \lambda }\left(a - g^{\theta_6}[ \theta_4 + \theta_5 A_1^{\theta}(t) ]  \right)^2  \right\} , \]
where $A^{\theta}$ is parameterized by a set of {\it different} parameters $(\theta_0,\theta_1,\theta_2,\theta_3)$ but in the {\it same} form as  $A_1^{\psi}$ in \eqref{eq:parametrize value function}. In total, $\theta\in \mathbb{R}^7$ consists of entries $\theta_0,\theta_1,\cdots,\theta_6$.

An alternative way is to engage neural networks. We can parameterize the value function by
\[ V^{\psi}(t,w,g) = \frac{w^{1-\gamma} \exp\{ (T-t) NN^{\psi}(t,g) \}}{1 - \gamma}- \frac{1}{1-\gamma}, \]
and  the policy by
\[ \bm\pi^{\theta}(a|t,g) = \frac{1}{\sqrt{ \frac{2\pi\lambda}{\gamma g}}}\exp\left\{ -\frac{\gamma g}{2 \lambda }\left(a - NN^{\theta}(t,g)\right)^2  \right\} , \]
where $NN^{\psi}$ and $NN^{\theta}$ are two neural networks with suitable dimensions of $\psi$ and $\theta$. Note these neural network constructions have also taken advantage of the theoretical results. %We have
% \[ \log\bm\pi^{\theta}(a|t,g) =-\frac{\gamma g}{2 \lambda }\big(a - NN^{\theta}(t,g)\big)^2 - \frac{1}{2}\log \frac{2\pi \lambda}{\gamma g } , \]
% \[ \frac{\partial }{\partial \theta}\log\bm\pi^{\theta}(a|t,g) = \frac{\gamma g}{ \lambda }\big(a - NN^{\theta}(t,g)\big) \frac{\partial NN^{\theta}(t,g)}{\partial \theta}.  \]

Finally, we use the stochastic approximation algorithm to search for the root to the estimating equations \eqref{eq:rl td equation general} with the test functions chosen as \eqref{eq:choice of test function}, where all the processes and integral are approximated via discretization in a way similar to that described in Subsection \ref{BS}.  We summarize these procedures as Algorithms \ref{algo:online incremental} and \ref{algo:offline incremental}  in both online and offline settings in Appendix \ref{appendix:algorithm}.
%\begin{equation}
%\label{eq:rl td equation}
%\begin{aligned}
%& \E\left[\int_0^T \xi_t \frac{ \dd \hat V^{\psi}(t,W_t^{\hat{\bm\pi}^{\theta}}, G_t)}{(1-\gamma) \hat V^{\psi}(t, W_t^{\hat{\bm\pi}^{\theta}}, G_t) + 1}  \right] = 0, \\
%& \E\left[ \int_0^T \eta_t \frac{\partial }{\partial \theta}\log\bm\pi^{\theta}(a_t^{\bm\pi^{\theta}}|t, G_t) \frac{ \dd \hat V^{\psi}(t,W_t^{\hat{\bm\pi}^{\theta}}, G_t)}{(1-\gamma) \hat V^{\psi}(t, W_t^{\hat{\bm\pi}^{\theta}}, G_t) + 1}  \right] = 0,
%\end{aligned}
%\end{equation}
%for suitably chosen ``test processes'' $\xi_t,\eta_t$ that are adapted to the filtration generated by $\{W_t^{\bm\pi^{\theta}}, G_t \}$. The temporal--difference (TD) type of online and offline algorithms use stochastic approximation to solve \eqref{eq:rl td equation} and  obtain $(\psi,\theta)$, which are summarized in the following Algorithms \ref{algo:online incremental} and \ref{algo:offline incremental}. Therein, we choose the most common and the simplest test processes used in literature: $\xi_t = \frac{\partial \hat V^{\psi}}{\partial \psi}(t,W_t^{\hat{\bm\pi}^{\theta}},G_t)$, and $\eta_t = 1$.

\section{Numerical Studies}
\label{numerical}
\subsection{Simulation with Synthetic Data}
A key advantage of a simulation study is that we have the ground truth (``omniscient") solutions available to compare against the learning results. In this subsection we report our numerical study with synthetic data, where sample paths of stock price and instantaneous variance process are simulated using the Euler-Maruyama scheme. {The data are generated from the ``3/2 model" with $\delta = \mu-r$ and $\alpha=-1$. In this case, the stock price and factor dynamics are
\[ \frac{\dd S_t}{S_t} = \mu\dd t + \frac{1}{\sqrt{X_t}}\dd B_t,\ \ \ \; \dd X_t = \iota(\bar{x} - X_t)\dd t + \bar{\nu}\sqrt{X_t}(\rho \dd B_t + \sqrt{1-\rho^2}\dd \tilde{B}_t ). \]
It is a typical non-affine stochastic volatility model proposed by \citet{drimus2012options}. In the classical case ($\lambda = 0$), the optimal policy is given by $\bm u^*(t,x) = (\mu-r)x/\gamma + \rho \bar{\nu}A(t)x/\gamma$.
%since the instantaneous variance of stock return is $\frac{1}{X_t}$ that satisfies:
%\[ \dd (\frac{1}{X_t}) = \frac{\iota\bar{x} - \bar{\nu}^2}{X_t}( \frac{\iota}{\iota\bar{x} - \bar{\nu}^2}  - \frac{1}{X_t})   \dd t - \bar{\nu}(\frac{1}{X_t})^{\frac{3}{2}}(\rho \dd B_t + \sqrt{1-\rho^2}\dd \tilde{B}_t) .\]

% \eqref{eq:hjb 4} now becomes
% \begin{equation}
% \label{eq:hjb sv alpha -1}
% \begin{aligned}
% u_t & + \big( (1-\gamma)r - \beta \big) + \iota(\bar{x} - x)u_x + \frac{1}{2}\bar{\nu}^2 x(u_{xx} + u_x^2)  \\
% & + \frac{(1-\gamma)x}{2\gamma}\Big[ \big( \mu - r \big)^2 + 2\rho\big( \mu-r \big)\bar{\nu}u_x  + \rho^2\bar{\nu}^2u_x^2\Big]  = 0 .
% \end{aligned}
% \end{equation}

% \begin{equation}
% \label{eq:classical solution ode}
% \begin{aligned}
% & A' - \iota A + \frac{1}{2}\bar{\nu}^2A^2 + \frac{1-\gamma}{2\gamma}\big[ (\mu-r)^2 + 2\rho(\mu-r)\bar{\nu}A + \rho^2\bar{\nu}^2A^2 \big] = 0,\ A(T) = 0\\
% & B' + (1-\gamma)r - \beta + \iota \bar{x}A = 0,\ B(T) = 0.
% \end{aligned}
% \end{equation}

The parameters are modified from the estimated values in \cite{chacko2005dynamic}, namely, $\delta = 0.2811$, $r = 0.02$, $\alpha=-1$, $\iota = 0.1374$, $\bar{x} = 35$, $\bar{\nu} = 0.9503$, and $\rho = 0.5241$.}\footnote{Under the originally estimated parameters in \cite{chacko2005dynamic}, the buy-and-hold is almost the optimal policy. To avoid this coincidence, we modify some parameters so that  different methods produce distinct results. } The risk aversion coefficient is taken as $\gamma=3$, which is a common value estimated from the aggregated growth and consumption data \citep{kydland1982time}. We further set the investment horizon $T=1$ (year), the initial wealth $w_0 = 1$, initial market factor $x_0 = \bar x$,  the temperature  parameter $\lambda = 0.1$, and the time discretization step size $\Delta t = \frac{1}{250}$. To mimic a real-world scenario, we generate a training dataset with daily data for  20 years, and each time we randomly sample a consecutive subsequence from that dataset with a length of 1 year as one episode for training (i.e. for updating the parameters $(\psi,\theta)$). The batch size for training is set to $m_{train}=16$. The initial learning rate is set to be 0.01 and decays as $l(j) = j^{-1/2}$. In total we carry out 2000 episodes for learning. On the other hand, the test set contains $N_{test}=10^4$ independent wealth trajectories, each generated from an episode having one-year length under the (deterministic) mean policy with the parameter $\theta$ learned from the training. %We then use these $10^4$ samples to calculate the average utility.
We reiterate that, in view of  Theorem \ref{thm:hjb solution}, randomized policies are used for training, while the mean of the learned randomized policy is used for testing.

%We further add noise to the observed volatility by $\tilde{G}_t = (\sqrt{G_t} + \epsilon_t)^2$, where $\epsilon_t\sim \mathcal{N}(0,0.002^2)$. Then the estimated parameters become $\hat{\bar{\nu}} = 0.0557$, $\hat{rho} = -0.264$, $\hat{\bar{x}} = 0.233$, $\hat{delta} = 1.284$, $\hat{\iota} = 1.533$, and $\hat{\alpha}=0.482$.

%Parameters in our algorithm are initialized by $\psi = 0.1$, $\theta = 0$.

%In our experiment, we always report two aspects of our algorithm. First we report the expected payoff on the test set to illustrate the convergence and final outcome of our methods. Second, we compute the average squared difference between the learned action and the theoretically optimal one, called ``misallocation''. In addition, we

For the simulation study we apply throughout the offline algorithm, Algorithm \ref{algo:offline incremental}, for learning/training. Moreover,
we implement two versions of function approximation for execution. One uses the specific parametric forms  motivated by the theoretical  solutions, denoted by ``This Paper - Specific Form''. The other one applies neural networks, denoted  by ``This Paper - Neural Network''. In particular, for the latter  we use two three-layer neural networks to approximate the value function and the randomized policy, respectively. We then compare these algorithms with the ground truth (``Omniscient") as well as two other methods. The first one is a na\"ive buy-and-hold policy (``B-H") that only holds the risky asset throughout without rebalance. It can also be regarded as the benchmark for investment if the risky asset is a market index (e.g. S\&P 500). The second is an estimate-and-plug-in policy based on the stochastic volatility model (``Est-SV") with the analytical solutions given by Lemma \ref{lemma:classical solution}. We employ a maximum likelihood estimation approach to estimate the parameters of the 3/2 model  using the  training set (with the length of 20 years).\footnote{The estimation is carried out by maximizing the likelihood function by the gradient ascent algorithm. The log-likelihood function of the data is approximated based on the Euler-Maruyama discretization of the SDEs, which coincides with the actual data generation process used in the simulation study.} We also implement the empirical risk minimization (ERM) approach using the same policy class approximations, denoted as ``ERM - Specific Form" and ``ERM - Neural Network", respectively. We emphasize that all the methods are tested on the same 20-year simulated data set and differ only in the loss function used during training and the parameters to be learned. %The estimated parameters are $\hat{\delta} = 0.403$, $\hat{\alpha}=-0.808$, $\hat{\iota} = 0.052$, $\hat{\bar{x}} = 9.621$, $\hat{\bar{\nu}} = 0.567$, and $\hat{\rho} = 0.531$.\footnote{Compared with the true values of these parameters, we can see that the volatility and correlation are estimated reasonably well, while  the drift parameters,  including the risk premium and the level and speed of mean-revision, are estimated poorly. Similar observations are well documented in literature.}

We use two performance criteria to compare the different methods. The first one is the average utility value on the test set. Specifically, given a deterministic policy obtained from training under a given method, we apply it to the test set and obtain $N_{test}$ independent one-year wealth trajectories. Denote the terminal wealth of these trajectories by $\hat{W}_T^{(i)}$, $i=1,2,\cdots,N_{test}$. Then the average utility is  $\frac{1}{N_{test}}\sum_{i=1}^{N_{test}} \frac{ \hat{W}_T^{(i)^{1-\gamma}}-1}{1-\gamma} $.
The second criterion is the equivalent relative wealth loss,  computed by finding $\Delta$ such that $V^{(0)}\left( 0, w_0(1 - \Delta),x_0 \right) = \text{average payoff on the test set}$, where $V^{(0)}$ is the optimal value function of the classical Merton problem under the true model.

Finally, to examine statistical significance of the proposed methods, we repeat the above simulation runs for 100 times with different random seeds. That is, for every simulation run, we first generate training data with a 20-year length and then apply each method to the same training data. After having obtained a learned/estimated policy through training, we calculate  its two performance criteria on the same testing data, which consists of 10000 independent 1-year trajectories. The results, including both the averages and standard errors of these 100 simulation runs, are summarized in the upper panel of Table \ref{tab:summary numerical repeated}. The B-H policy is independent of any model or learning specifications, yielding about 2/3 of the omniscient utility value and 18.28\% loss of initial endowment. On average, with the correctly specified model class, the Est-SV policy performs much better than B-H, generating 95.37\% of the optimal utility and 2.96\% loss in wealth. The RL algorithm with the specific parametric form  outperforms B-H and Est-SV by considerable margins, with very small losses of utility (97.03\% of the optimal utility) and relative wealth (2.16\%). By contrast, the RL with neural networks performs relatively worse, roughly on par with Est-SV. We have done extensive experiments and observed that this finding is robust with respect to the structures of the neural networks used. The reason behind the discrepancy between the two RL methods is that in the simulation study, the specific parametric method uses the {\it correct} form of the optimal policy that corresponds to the true underlying data-generating process, while neural networks do not use much such structural information. In the experiment presented here, the size of the training dataset is relatively small; so approximation with the correct form performs  better than using general neural networks, the latter likely over-fitted. Indeed, the training set contains only 20-year data so the distribution of the training set may considerably differ from the theoretical distribution due to sampling errors. Moreover, as we take 2000 episodes for training, the data in those episodes overlap and are hence not mutually independent. To this point, we provide extra numerical results in Appendix \ref{appendix:additional numerical results} based on a huge amount of data, where new and {\it independent} trajectories are generated in each training episode. In that experiment, neural networks perform equally well as the other RL method. ERM performs the worst among the methods given the same size of training data, due to the effects identified in \citet{reppen2023deep}. Its underperformance in such a small-data regime is  also consistent with the results under the Black-Scholes market shown in Figure \ref{fig:gbm lambda}.

\begin{table}[htbp]
\centering
\caption{\textbf{Performance comparison of different methods under 100 simulation runs.} We compute the average utility value under each policy based on independent $10^4$ one-year wealth trajectories, and then use the formula in Definition \ref{def:erwl} to convert average utility to equivalent relative wealth loss. B-H stands for the buy-and-hold policy, and Est-SV for the estimate-and-plug-in policy. ``ERM - Specific Form" and ``ERM - Neural Network" are two empirical risk minimization variants with the same parameterization as ours. Each policy other than Omniscient and B-H is obtained from a simulated training set of daily data for 20 years, and the simulation is repeated with 100 independent runs. The numbers in the bracket indicate the standard errors.}
\begin{tabular}{cccc}
\toprule
Volatility	&	Method	&  Utility &  Equivalent Relative Wealth Loss  \\
\midrule
& Omniscient & 0.303 & 0 \\
& B-H  & 0.201 & 18.28\% \\
\midrule
Exact	&		This Paper - Specific  Form & {\bf 0.294}  & {\bf 2.16}\% \\
&		& (0.001) & (0.24\%) \\
&		This Paper - Neural Network  & 0.290 &  3.04\% \\
&		& (0.001) & (0.21\%) \\

&		Est-SV  & 0.289 & 2.96\% \\
&		& (0.002) & (0.35\%) \\
&		ERM - Specific Form  & 0.233 & 13.91\% \\
&		& (0.002) & (0.28\%) \\

&		ERM - Neural Network  & 0.236 & 13.35\% \\
&		& (0.002) & (0.30\%) \\
& & & \\		
Noisy &		This Paper - Specific  Form & 0.286  & 3.84\% \\
&		& (0.002) & (0.32\%) \\
&		This Paper - Neural Network  & {\bf 0.290} &  {\bf 3.04}\% \\
&		& (0.001) & (0.21\%) \\

&		Est-SV  & 0.238 & 28.01\% \\
&		& (0.005) & (0.21\%) \\

&		ERM - Specific Form  & 0.233 & 13.93\% \\
&		& (0.002) & (0.28\%) \\

&		ERM - Neural Network  & 0.236 & 13.35\% \\
&		& (0.002) & (0.30\%) \\
\bottomrule
\end{tabular}%
\label{tab:summary numerical repeated}%
\end{table}%

Next we examine the robustness of our algorithms with respect to the observable volatility process, motivated by the considerations that in practice one only has access to an approximated value of the volatility, and/or that the stochastic volatility model is wrongly specified. To this end, we construct a noisy observation $\tilde G_t = (\sqrt{G_t} + 0.02\xi_t)^2$,  where $\xi_t\sim \mathcal{N}(0,1)$ are i.i.d. at (daily) observation times. This construction applies to both the training and testing datasets. This implies that the observed volatility signals deviate from the true one by 2\% on average, and the agent only observes $(S_t,\tilde G_t)$.
The corresponding comparisons across various methods are presented in the lower panel of Table \ref{tab:summary numerical repeated}.
Compared with the previous results, the specific parametrization RL method still performs well and is only slightly worse than the case with exact volatility, while the neural network-based method yields almost identical performance to its non-noisy counterpart.
Note that B-H does not rely on volatility; so it has identical results as before. ERM also performs similarly. It is most noteworthy, however, that the performance of Est-SV drops dramatically owing to the contaminated data, which once again confirms the sensitivity (and drawbacks) of the conventional plug-in methods. By contrast, our RL methods are ``model-parameter-free'' and learn policies directly, resulting in a much more robust performance.

\subsection{Empirical Study with Real Market Data}
We study dynamic allocation between the S\&P 500 index  and a money market account with $r=2\%$ risk-free interest rate to illustrate the performance of our RL algorithms in the real market. %We assume there is no transaction cost and no leverage or borrowing constraints.
S\&P 500 is one of the most actively traded indices and its option market is also highly liquid. Therefore, we can easily obtain volatility-related data from the market. In particular, VIX is an index administered by CBOE (Chicago Board Options Exchange) since 1990 based on  option prices that reflects the market-priced average forward-looking volatility of the S\&P 500 index, and is widely considered to be a proxy of  the instantaneous volatility. VIX itself is a traded future with options written on it. In our empirical study, we take the S\&P 500 index as the risky asset and VIX as a proxy for its volatility, both observable.  We use data from 1990-01-01 to 2025-11-15, employing the first 10 years (up to 1999-12-31) as the pre-training period and the remainder as the testing period. During the former period, we apply our offline algorithm, Algorithm \ref{algo:offline incremental}, to learn the parameters $(\psi,\theta)$ and set the learned ones as the initial parameters for the latter period. Then we use the online algorithm, Algorithm \ref{algo:online incremental}, to learn and implement optimal  Merton's strategies as we go.
We fix our initial wealth on 2000-01-01 to be 1 dollar and take the risk aversion parameter as $\gamma = 3$. %\footnote{With typical returns and volatility of S\&P 500 index, choosing small risk aversion parameter implies large leverage ratio which is not realistic. $\gamma = 3$ is a reasonable value that generates reasonable leverage ratio. The leverage ratio is defined to be the proportion allocated on the risky asset.}
The  benchmark policies for comparision are still  the buy-and-hold  (B-H), the estimate-and-plug-in (Est-SV), and ERM. We do not allow leverage or borrowing for all the policies under comparison; therefore, if a method suggests taking leverage or short selling, we truncate the portfolio value to lie within the interval $[0,1]$.

Also, to avoid seasonality that depends on the investment horizon, we consider only  time-invariant policies, which can be viewed as the limit when the time-to-maturity approaches infinity. This seems reasonable given that we have a rather long testing period.  Note that for a stochastic volatility model, such time-invariant policies still result in time-variant portfolios via the (time-variant) instantaneous volatility. The form of the optimal policy in \eqref{eq: explore optimal solution asymptotic lambda general} then becomes
\[ \operatorname{Mean}(\bm\pi^*(g)) = C_1 g^{C_2} ,\;\;\; \operatorname{Var}(\bm\pi^*(g)) = \frac{\lambda}{\gamma g}\]
for some constants $C_1, C_2$.

ERM is implemented by minimizing the empirical risk function using the same policy class parameterizations as in our approach. After being pre-trained on the first 10 years of data (up to 1999-12-31), the policy class is re-trained monthly starting from 2000, using the most up-to-date data. In other words, the learned policy is applied out-of-sample for one-month periods with daily rebalancing frequency. We employ two different parameterization methods, as in our approach: one with a specific parametric form and the other with neural networks. Interestingly, due to the effect of truncation, ERM with neural networks consistently suggests ``all-in on the index". As a consequence, it coincides with the B-H benchmark; so we only report results for ERM with the specific parametric form. The Est-SV is implemented as follows. %Est-GBM assumes a geometric Brownian motion model so the implied policy is time-invariant. We use the rolling window of the length 10 years to estimate the expected return and its variance, and then plug-in to the form of the optimal solution under the GBM model.
First, we also restrict to time-invariant policies. We use a rolling window with a length of 10 years to estimate the model parameters and then plug-in to the analytical form of the optimal solution under the SV model. To save computational cost and avoid re-estimating the whole model every day, we only update the estimation of model parameters by maximizing the log-likelihood function along the gradient ascend direction for one step during the testing period.%\footnote{This is analogous to the online updating  in RL algorithms. The construction of the log-likelihood function involved is described in Footnote 14.}

A comparison of different methods  is summarized in Table \ref{tab:real data},  in terms of several commonly used metrics including (annualized) return, volatility, Sharpe ratio, (downside) semi-volatility, Sortino ratio, Calmar ratio, maximum drawdown, and recovery time.
Among them Sharpe ratio is the most important and popular criterion because the essential goal of the Merton problem is to maximize the risk-adjusted return.
We observe the two RL methods outperform the other two methods in all the metrics except the annualized return (B-H has 5.9\%, slightly higher than 5.6\% by RL with neural networks).
In particular, RL with neural networks beats the other methods by significant margins in most criteria including the Sharpe ratio. Moreover,  the two RL methods have remarkably smaller maximum drawdowns during the whole period in which the market experienced a 56.8\% drawdown.  Even more notably,  their recovery times are decisively and overwhelmingly shorter. These observations  indicate that RL strategies not only perform strongly but also robustly, and  react to the environment change and pivot very quickly.

%We see that although RL with neural networks has a slightly lower annualized return than B-H, it
%achieves the highest risk-adjusted return (the Sharpe ratio), followed by B-H and  RL with specific form. Notably, both RL methods have significantly smaller maximum drawdowns and even more significantly quicker recovery times compared with B-H, the market index. This shows that the RL strategies are remarkably more stable and robust. % cannot outperform the naive B-H. The RL approach with neural networks, however, show promising results that produce slightly lower return but much higher risk-adjusted performance. Why empirical results favor the neural networks? One possible reason is that the particular parametric form we adopted is still rationalized by a stochastic volatility model, which may not be valid in the real market. In this case, the flexible structure of neural networks help to identify other possible forms of investment strategies to make use of the VIX signal.

\begin{table}[htbp]
\centering
\caption{\textbf{Comparison of out-of-sample performances of different methods from January 2000 to November 2025.} We report the (annualized) return (Rtn), volatility (Vol), Sharpe ratio, (downside) semi-volatility (Semi-Vol), Sortino ratio, Calmar ratio, maximum drawdown (MDD), and recovery time (RT). The risk-free interest  $r = 0.02$.}
\begin{tabular}{lcccccccc}
\toprule
Method & Rtn   & Vol   & Sharpe & Semi-Vol & Sortino & Calmar & MDD    & RT \\
\midrule

This Paper: Specific & 0.037 & \textbf{0.078} & 0.221 & \textbf{0.060} & 0.289 & 0.069 & \textbf{0.251} & 282 \\
This Paper: Neural Network & 0.056 & 0.118 & \textbf{0.306} & 0.088 & \textbf{0.411} & \textbf{0.107} & 0.339 & \textbf{202} \\
B-H   & \textbf{0.059} & 0.194 & 0.203 & 0.142 & 0.277 & 0.069 & 0.568 & 1376 \\
%Est-GBM & 0.035 & 0.139 & 0.109 & 0.102 & 0.148 & 0.032 & 0.478 & 4097 \\
Est-SV & 0.028 & 0.100 & 0.075 & 0.075 & 0.101 & 0.017 & 0.440 & 4248 \\
ERM   & 0.055 & 0.189 & 0.186 & 0.140 & 0.252 & 0.058 & 0.601 & 1463 \\

\bottomrule
\end{tabular}%
\label{tab:real data}%
\end{table}%

\begin{figure}[h]
\centering
\includegraphics[width=0.5\textwidth]{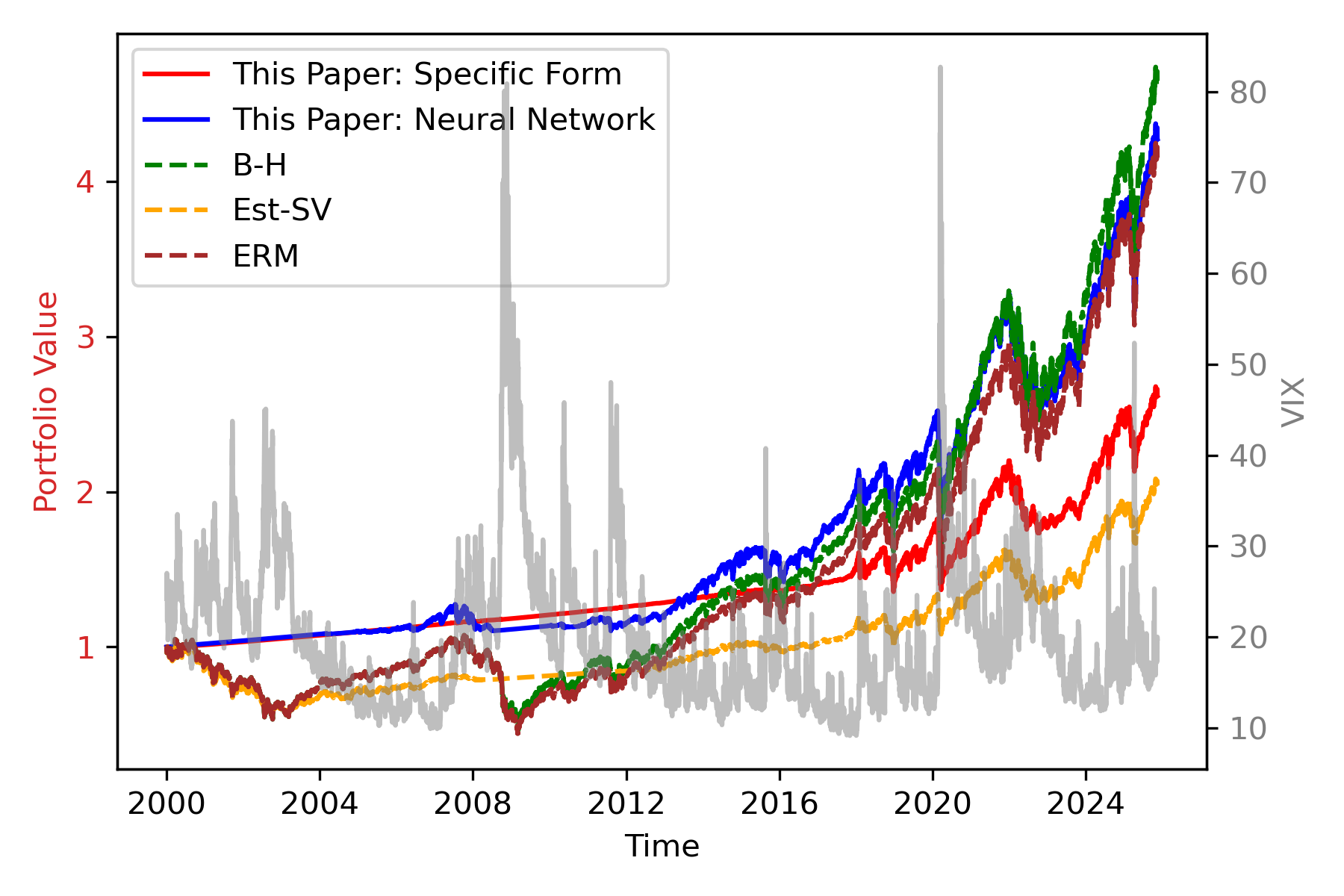}
\caption{\textbf{Wealth trajectories  of portfolios under different policies.} The gray plot is the VIX index whose vertical axis is on the right. The other plots are the trajectories of the portfolio values under different methods and are all normalized to 1 initially.}
\label{fig:real data wealth}
\end{figure}

While Table \ref{tab:real data} gives a glance of overall and average performance comparison over 25 years, we now inspect the wealth trajectories under different policies, presented in
Figure \ref{fig:real data wealth}. It is clear that RL with neural networks outperforms (in terms of portfolio worth) all the others prior to around 2020, taken over by B-H only after 2020. However, both RL portfolios are much less volatile than B-H, corroborating the findings of Table \ref{tab:real data}. In particular, the RL strategies considerably  and consistently beat the other two during the first 10 years, 2000-2010. Recall that this is an extremely volatile period, including  two bear markets, the dot com bubble burst in the early 2000s and the  financial crisis during 2007-2008.\footnote{The robustness, especially the outperformance during bear markets and the significantly shorter recovery time, of RL strategies devised from continuous-time theory is also documented for mean-variance portfolio choice; see \citet{huang2022achieving,huang2024mean}.}

\begin{figure}[h]
\centering
\includegraphics[width=0.5\textwidth]{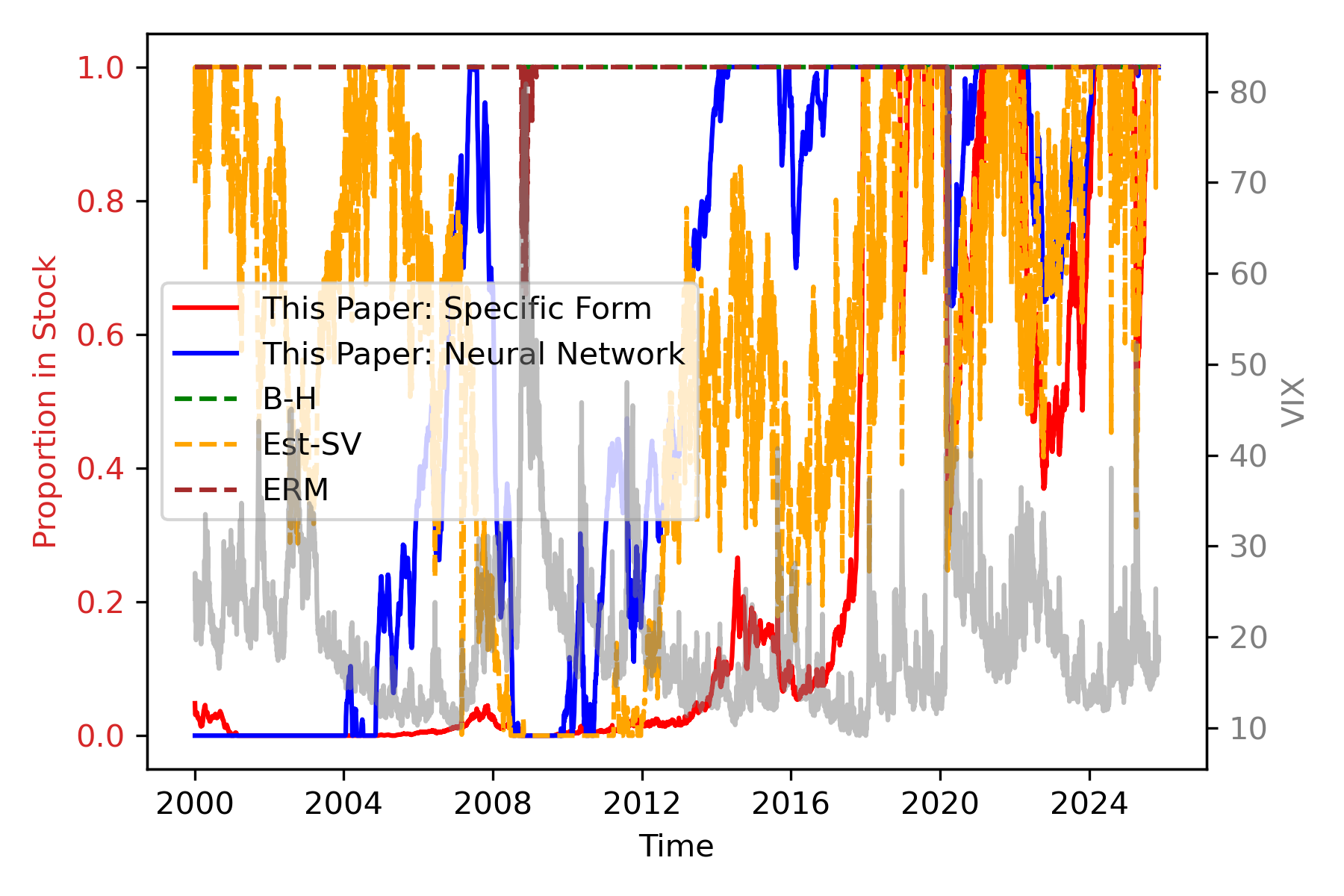}
\caption{\textbf{Trajectories  of risky proportions  under different policies.}  In our study, proportions invested in S\&P 500 are restricted to be between 0 and 1. The gray curve is the VIX index whose vertical axis is on the right. The other curves are the trajectories of the proportions of the risky investment under different methods.  The initial allocations of all the methods (except B-H) are based on the pre-training period from 1990 to 1999.}
\label{fig:real data action}
\end{figure}

We further examine the proportions of wealth invested in the risky asset under different strategies, depicted in  Figure \ref{fig:real data action}. An interesting observation is that in the first half of the 2000-2010 overall bear period, the two RL-portfolios, especially the one with  specific form, do not hold much risky asset, unlike Est-SV. %This is largely due to the offline pretraining results from 1990 to 1999.
It demonstrates how the RL approach fundamentally differs from the traditional plug-in approach: Est-SV estimates model parameters statistically based on the market data in the previous 10 years (1990-1999) that had a positive risk premium and that were characteristically different from those in the early 2000s. By contrast, RL learns portfolio strategies through real-time interactions with the market and pivots
timely to more conservative ones after the market pivots. On the other hand,  all the methods detect buying signals after 2010, with RL with neural networks being the first to react and gradually start overweighting the risky asset. Finally, ERM is almost identical to B-H (except during the 2007-2008 financial crisis) maintaining an all-in position most of the time, which indicates a slower adaptation to new information.

The empirical results decisively indicate that the RL methods outperform the others on all fronts. Regarding the competition between the two RL algorithms, the one using neural networks outperforms the other by a significant margin, contrary to the results from the simulation study. This discrepancy arises because the specific parametric form we adopt follows from a specific 3/2 model, which is almost certainly invalid in real markets, while the flexible structure of neural networks allows for the identification of alternative investment strategies by fully exploiting the VIX signals.

\section{Conclusions}
\label{sec:concl}
RL, as one of the cutting-edge technologies in artificial intelligence, has been applied to various fields. The central component of RL is exploration, which is carried out by policy randomization to broaden the action space aiming at understanding the interactions between an unknown  environment and actions for improving and optimizing decision-making.

Applications of RL in finance however, especially in portfolio choice, is still in the early innings. One question is that, unlike  the bandits problem, stock data are exogenous and hence there is no need to ``explore"---to actually try different portfolios to see the outcomes---if the market impact is ignored. In other words, there is no exploration-exploitation tradeoff because no extra information is gained by trial and error.

In this paper, we argue otherwise. To wit, we show that RL including policy randomization  can go beyond the conceptual role of exploration; it can actually be used also as a {\it technical} tool to solve a ``model-free" problem that otherwise cannot be solved satisfactorily by the conventional model-based methods. We do this in the setting of Merton's investment problem in an incomplete market and derive its data-driven solutions. More precisely, we demonstrate that,  in spite of having no informational benefit, RL can still be used to learn optimal portfolio policies in a model-free manner by employing randomized actions. We propose an auxiliary  relaxed control problem with a special class of Gaussian policies within the continuous-time RL exploratory framework developed by \citet{wang2018exploration} and show that the optimal solution of this auxiliary problem gives rise to  that of the original  Merton problem.
A key insight is that  exploration-exploitation tradeoff in the current setting of a small investor is not about information gains versus payoff losses, but about the strength of the learning signals (the gradient estimates of  the objective function) versus their reliability (the variance of the gradient estimates). It goes without saying that the RL approach can be extended readily to the problem with a large investor in which the RL will play both the conceptual and technical roles.
As such, we believe that the paper  resolves the long-standing question about the necessity and applicability of RL in portfolio choice.

We develop  an actor-critic RL algorithm for learning optimal policies and value functions iteratively. Through policy evaluation and policy update, we show that such an iterative procedure yields monotonically improving policies. Using a stochastic volatility environment as an example, we explain why the traditional model-based, plug-in methods may fail due to sensitivity to model estimation errors. By contrast, the RL methods are model-free and learn optimal policies directly from data, which is naturally robust to the said estimation errors. Numerical results based on both synthetic and market data forcefully demonstrate the efficiency and robustness of our methods against traditional  plug-in methods.

This paper brings about many open research questions. A fascinating one is  to fully understand the general interactions between the randomness injected by randomized policies and the randomness in the market, as well as their joint impacts on learning performance.

\theendnotes

\bibliographystyle{informs2014}
\bibliography{ref}

\bigskip
\newpage
\begin{APPENDICES}

%\ECSwitch

%\ECDisclaimer
%%%%%%%%%%%%%%%%%%%%%%%%%%%%%%%%%%%%%%%%%%%%%%%%%%%%%%%%%%

%%% Main head for the e-companion
\ECHead{Electronic Companion to ``Data-Driven Merton's Strategies via Policy Randomization''}

\section{Motivation of Formulation  \eqref{controlled_system}}\label{appendix:motivation}
We explain the exploratory formulation \eqref{controlled_system} by starting with a discrete-time setting for ease of understanding. Divide the whole time interval $[0,T]$ into small intervals of size $\Delta t$. Let $R_t:=\log W_t$ be the wealth log-return.  Given an action $a \in \mathbb R$, the instantaneous change of the log return process in the interval $[t,t+\Delta t]$ is
$$
\Delta R_t=\left[r+(\mu_t-r)a-\frac{1}{2}\sigma_t^2a^2 \right]\Delta t +\sigma_t a \Delta B_t.
$$	
Now, we assume that the agent takes action  randomly according to a policy distribution $\pi_t$ that is independent of the underlying Brownian motions in the market. %This implies that the action $u$ should be a random variable with mean and variance given by $\pi_t$.
Focusing on the first and second moments of the randomized policy, we replace $a$ with $e_t+v_t\varepsilon_t$, where $\varepsilon_t$ is a random variable with zero mean and unit variance independent of $B_t$ and $\tilde{B}_t$,
$$e_t=\int_{\mathbb{R}} a \bm\pi_t(a)\dd a, \ \ \mbox{and } \ \  v_t=\sqrt{\int_{\mathbb{R}} a^2 \bm\pi_t(a)\dd a-(\int_{\mathbb{R}} a \bm\pi_t(a)\dd a)^2 }.$$
It follows that %can rewrite the instantaneous change as
\begin{eqnarray*}
\Delta R_t&=&\left[r+(\mu_t-r)(e_t+v_t \varepsilon_t)-\frac{1}{2}\sigma_t^2(e_t+v_t \varepsilon_t)^2 \right]\Delta t +\sigma_t (e_t+v_t \varepsilon_t) \Delta B_t \\
&=&\left[r+(\mu_t-r)e_t-\frac{1}{2}\sigma_t^2(e^2_t+v_t^2) \right]\Delta t +\sigma_t e_t \Delta B_t+\sigma_t v_t \varepsilon_t \Delta B_t+ \text{Residual}_t,
\end{eqnarray*}
where the residual term $\text{Residual}_t$ is given as follows:
$$
\text{Residual}_t=(\mu_t-r)v_t \varepsilon_t \Delta t- \sigma_t^2 e_tv_t \varepsilon_t \Delta t -\sigma_t^2 v^2_t(\varepsilon^2_t-1)\Delta t.
$$
{Since the residual term is a mean zero random variable of size $O(\Delta t)$ and the strategy noises $\varepsilon_t$ are mutually independent between time intervals, by the law of large numbers, the residual term will vanish when we take the sum over the whole time interval and send $\Delta t$ to zero. %On the other hand,
In addition, as $\varepsilon_t \Delta B_t$ is a mean zero random variable of size $O(\sqrt{\Delta t})$, its summation is asymptotically Gaussian by the central limit theorem.} Furthermore, we have  $\text{Cov}(\varepsilon_t \Delta B_t,\Delta B_t)=0$ and $\text{Cov}(\varepsilon_t \Delta B_t,\Delta \tilde{B}_t)=0$ as $\varepsilon_t$ is independent of $B_t$ and $\tilde{B}_t$. Thus, $\varepsilon_t \Delta B_t$ can be approximately treated as the increment of another Brownian motion independent of $B_t$ and $\tilde{B}_t$.
It is not hard to verify that
\begin{eqnarray*}
\mathbb E\left[ \Delta R_t\right]&=& \left[r+(\mu_t-r)e_t-\frac{1}{2}\sigma_t^2(e^2_t+v_t^2) \right]\Delta t\\
&=&
\left[r+(\mu-r)\int_{\mathbb{R}} u\bm\pi_t(u)\dd u-\frac{1}{2}\sigma_t^2\int_{\mathbb{R}} u^2 \bm\pi_t(u)\dd u \right]\Delta t,\\
\text{Var}\left[ \Delta R_t \right]&=&\sigma_t^2 (e_t^2+v_t^2)\Delta t+o(\Delta t)=\sigma_t^2 \int_{\mathbb{R}} u^2 \bm\pi_t(u)\dd u \Delta t+o(\Delta t),\\
\text{Cov}\left[ \Delta R_t,\Delta X_t \right]&=&\rho \nu \sigma_t e_t \Delta t+o(\Delta t)=\rho \nu \sigma_t \int_{\mathbb{R}} u \bm\pi_t(u)\dd u \Delta t +o(\Delta t).
\end{eqnarray*}
This suggests that at the continuous-time limit, $R$ satisfies the following SDE
\begin{eqnarray*}
\dd R_t=&\left[r+(\mu_t-r)\operatorname{Mean}(\bm\pi_t)-\frac{1}{2}\sigma_t^2
(\operatorname{Mean}(\bm\pi_t))^2
-\frac{1}{2}\sigma_t^2\operatorname{Var}(\bm\pi_t)\right]\dd t\\
&\;\;+\sigma_t \left[ \operatorname{Mean}(\bm\pi_t) \dd B_t+\sqrt{ \operatorname{Var}(\bm\pi_t) }\dd \bar{B}_t\right],\ \ R_0 = \log w_0,
\end{eqnarray*}
where $\bar B_t$ is another Brownian motion that is  mutually independent of $B_t$ and $\tilde B_t$.  As discussed earlier,  $\bar B_t$ is introduced to model the additional noise caused by policy randomization and can be regarded as a ``random number generator" to generate a randomized policy. The coefficient of the $\dd\bar B_t$ term involves the variance of $\bm\pi_t$, measuring how much additional noise is introduced into the system.

Applying It\^o's formula to the above equation we get that $W^{\bm\pi}_t=e^{R_t}$ satisfies
the exploratory dynamics \eqref{controlled_system}.
%	\begin{equation*}
%	\frac{\dd W^{\bm\pi}_t}{W^{\bm\pi}_t} = \left[r+(\mu_t-r)\operatorname{Mean}(\bm\pi_t)\right]\dd t +\sigma_t \left[ \operatorname{Mean}(\bm\pi_t) \dd B_t+\sqrt{ \operatorname{Var}(\bm\pi_t) }\dd \bar{B}_t\right],
%\end{equation*}
As indicated by the above analysis, this exploratory formulation captures the information up to the second order. \citet{jia2025accuracy} provide a rigorous proof of how the wealth processes under portfolios time-discretely sampled from $\bm\pi$ converge weakly to the solution of \eqref{controlled_system} when the time step goes to 0.

%\section{Expressions for the Solutions to ODEs}
%\label{appendix:expression}
%In \eqref{eq:classical solution ricatti},
%\[ \psi_0 = -\frac{\sqrt{\iota^2 \gamma - (1-\gamma) \delta\bar\nu(\delta\nu + 2\iota\rho)}}{\sqrt{\gamma}}, \ \psi_1 = \frac{(1-\gamma)\delta^2}{\bar{\nu}^2[\rho^2 + \gamma(1-\rho^2)]} ,\]
%\[ \psi_{2,3} = \frac{\iota \gamma - (1-\gamma)\delta\bar\nu\rho \pm \sqrt{\gamma} \sqrt{\iota^2 \gamma - (1-\gamma) \delta\bar\nu(\delta\nu + 2\iota\rho)} }{\bar{\nu}^2[\rho^2 + \gamma(1-\rho^2)]} ,\]
%\[ \psi_4 = (1-\gamma)r -\beta + \iota\bar x\psi_3,\ \psi_5 = \frac{2\gamma\iota\bar x}{\bar{\nu}^2[\rho^2 + \gamma(1-\rho^2)]}  .\]
%
%In \eqref{eq:explode ode},
%\[ \psi_0 = \frac{\iota \gamma - (1-\gamma)\delta\bar\nu\rho }{\bar{\nu}^2[\rho^2 + \gamma(1-\rho^2)]},\ \psi_1 = \frac{\sqrt{\gamma}\sqrt{-\iota^2 \gamma + (1-\gamma) \delta\bar\nu(\delta\nu + 2\iota\rho)}}{\bar{\nu}^2[\rho^2 + \gamma(1-\rho^2)]} ,\ \psi_2 = \frac{\sqrt{-\iota^2 \gamma + (1-\gamma) \delta\bar\nu(\delta\nu + 2\iota\rho)}}{2\sqrt{\gamma}},\]
%\[\psi_4 = (1-\gamma)r - \beta -\iota \bar{x} \psi_0,\ \psi_5 = -\frac{2\gamma\iota\bar{x}}{\bar{\nu}^2[\rho^2 + \gamma(1-\rho^2)]}.  \]

\section{Learning via Empirical Risk Minimization (ERM)}
\label{appendix:erm}
We document an alternative popular data-driven approach to  portfolio problems: optimizing policies  through ERM (see, e.g., \citealt{reppen2023deep}). Specifically, one parameterizes the portfolio policy by a deterministic and sufficiently smooth function of the factor: $a_t = \bm u^{\theta}(t,X_t)$, and  rewrite the wealth equation \eqref{classical_wealth} as
\[ \dd \log W_t^{\theta} = \bm u^{\theta}(t,X_t) \frac{\dd S_t}{S_t} + [1 - \bm u^{\theta}(t,X_t)] r \dd t - \frac{1}{2}\left( \bm u^{\theta}(t,X_t) \right)^2 \dd \langle \log S \rangle_t . \]
The derivative of $\log W_T^{\theta}$ in the parameter $\theta$ is
\[ \frac{\partial \log W_T^{\theta}}{\partial \theta} = \int_0^T  \frac{\partial \bm u^{\theta}}{\partial \theta}(t,X_t) \left( \frac{\dd S_t}{S_t} - r \dd t \right) - \left( \bm u^{\theta}\frac{\partial \bm u^{\theta}}{\partial \theta} \right)(t,X_t) \dd \langle \log S \rangle_t . \]
Therefore, the derivative of the objective in $\theta$ is
\[\begin{aligned}
& \frac{\partial }{\partial \theta}\E\left[ U\left( W_T^{\theta} \right) \right] = \frac{\partial }{\partial \theta}\E\left[\frac{\exp\{ (1-\gamma)\log W_T^{\theta}  \} - 1}{1-\gamma} \right] \\
= &  \E\left[ \left( W_T^{\theta} \right)^{1-\gamma} \frac{\partial \log W_T^{\theta}}{\partial \theta} \right] = \E\left[ \left( W_T^{\theta} \right)^{1-\gamma} \int_0^T  \frac{\partial \bm u^{\theta}}{\partial \theta}(t,X_t) \left( \frac{\dd S_t}{S_t} - r \dd t \right) - \left( \bm u^{\theta}\frac{\partial \bm u^{\theta}}{\partial \theta} \right)(t,X_t) \dd \langle \log S \rangle_t  \right] .
\end{aligned} \]

One then updates $\theta$ using a gradient-based algorithm. Note here we can calculate the gradient of the objective function due to the special structure of the wealth equation \eqref{classical_wealth} along with the assumption that the stock price $S_t$ and the market factor $X_t$ are both exogenous. In a more general setting, computing the gradient may require
the knowledge of model primitives.
%So, compared to the method employed in the main text of the present paper, this method is restricted to such special structures and  thus not a general methodology.
In addition, ERM cannot be applied in real-time (i.e. online)  because it requires the observation of the whole stock-factor-wealth process until $T$. %Moreover, the expression of the gradient is complicated, and one may experience higher variance when estimated by samples.

%Ultimately, they have a similar convergence rate, and the magnitude of the error remains close to that of the best performing proposed method. However, it is noticeable that it usually incurs larger errors when the learning is immature.

%\section{Numerical Algorithms}
\newpage
\section{Summary of Algorithms}
\label{appendix:algorithm}
\begin{algorithm}[!hbtp]
\caption{Online-Incremental Learning Algorithm}
\textbf{Inputs}: initial wealth $w_0$, initial stock price $s_0$, initial instantaneous variance $g_0$, horizon $T$, time step $\Delta t$, number of mesh grids $K$, initial learning rates $l_{\theta},l_{\psi}$ and learning rate schedule function $\ell(\cdot)$ (a function of the number of episodes), functional form  of parameterized  value function $\hat V^{\psi}(\cdot,\cdot,\cdot)$, functional form of parameterized policy function $\hat{\bm\pi}^{\theta}(a|t,g)$, interest rate $r$, risk aversion coefficient $\gamma$,  temperature parameter $\lambda$.

\textbf{Required program}: market simulator $(s',g') = \textit{Market}_{\Delta t}(t,s,g)$ that takes current time,  stock price, and instantaneous variance, $(t,s,g)$, as inputs and generates  stock price $s'$ and instantaneous variance $g'$ at time $t+\Delta t$ as outputs.

\textbf{Learning procedure}:
\begin{algorithmic}
	\STATE Initialize $\theta,\psi$.
	\FOR{episode $j=1$ \TO $\infty$} \STATE{Initialize $k = 0$. Observe initial wealth $w_0$, initial stock price $s_0$, and initial instantaneous variance $g_0$. Store $w_{t_k} \leftarrow  w_0$, $s_{t_k} \leftarrow  s_0$, $g_{t_k} \leftarrow  g_0$.
		\WHILE{$k < K$} \STATE{
			Generate action $a_{t_k}\sim \bm{\bm\pi}^{\psi}(\cdot|t_k,g_{t_k})$.
			
			Apply $a_{t_k}$ to market simulator $(s',g') = Market_{\Delta t}(t_k, s_{t_k}, g_{t_k})$, and observe new state $s',g'$. Store $s_{t_{k+1}} \leftarrow s'$, $g_{t_{k+1}} \leftarrow g'$.
			
			Compute  current wealth $w_{t_{k+1}} = w_{t_k} + w_{t_k}a_{t_k}\frac{s_{t_{k+1}}}{s_{t_k}} + w_{t_k}(1 - a_{t_k})r\Delta t $.
			
			Compute
			\[ \begin{aligned}
				\delta = & \frac{\hat V^{\psi}(t_{k+1},w_{t_{k+1}},g_{t_{k+1}}) - \hat V^{\psi}(t_{k},w_{t_{k}},g_{t_k})}{(1-\gamma) \hat V^{\psi}(t_{k},w_{t_{k}},g_{t_k}) + 1} .
			\end{aligned} \]

			Update $\theta$ and $\psi$ by
			\[ \psi \leftarrow \psi + \ell(j)l_{\psi} \delta \frac{\partial \hat V^{\psi}}{\partial \psi}(t_k,w_{t_k},g_{t_k}).  \]
			\[ \theta \leftarrow \theta + \ell(j)l_{\theta} \delta \frac{\partial }{\partial \theta}\log\hat{\bm\pi}^{\theta}(a_{t_k}|t_k,g_{t_k}) .\]
			Update $k \leftarrow k + 1$
		}
		\ENDWHILE	
		
	}
	\ENDFOR
\end{algorithmic}
\label{algo:online incremental}
\end{algorithm}

%We update $\psi$ by
%\[  \psi \leftarrow \psi + \alpha_{\psi} \frac{\partial V^{\psi}}{\partial \psi}(t,G_t,W_t)\left\{ \dd V^{\psi}_t + \lambda [(1-\gamma) V_t^{\psi} + \kappa_t][\frac{1}{2}\log\frac{2\bm\pi\lambda}{\gamma G_t} + \frac{1}{2}]\dd t - \beta V_t^{\psi}\dd t \right\} \]
%and update $\theta$ by
%\[ \theta \leftarrow \theta + \alpha_{\theta}  \frac{\gamma G_t}{ \lambda }\big(a - NN^{\theta}(t,G_t)\big) \frac{\partial NN^{\theta}}{\partial \theta}(t,G_t) \left\{ \dd V^{\psi}_t +\lambda [(1-\gamma) V_t^{\psi} + \kappa_t][\frac{1}{2}\log\frac{2\bm\pi\lambda}{\gamma G_t} + \frac{1}{2}]\dd t - \beta V_t^{\psi}\dd t \right\}  . \]
\begin{algorithm}[!hbtp]
\caption{Offline Learning Algorithm}
\textbf{Inputs}: initial wealth $w_0$, initial stock price $s_0$, initial instantaneous variance $g_0$, horizon $T$, time step $\Delta t$, number of mesh grids $K$, initial learning rates $l_{\theta},l_{\psi}$ and learning rate schedule function $\ell(\cdot)$ (a function of the number of episodes), functional form of parameterized  value function $\hat V^{\psi}(\cdot,\cdot,\cdot)$, functional form of parameterized policy function $\hat{\bm\pi}^{\theta}(a|t,g)$, interest rate $r$, risk aversion coefficient $\gamma$, temperature parameter $\lambda$.

\textbf{Required program}: market simulator $(s',g') = \textit{Market}_{\Delta t}(t,s,g,a)$ that takes current time,  stock price, and instantaneous variance $(t,x,g)$ as inputs and generates stock price $s'$ and instantaneous variance $g'$ at time $t+\Delta t$ as outputs.

\textbf{Learning procedure}:
\begin{algorithmic}
	\STATE Initialize $\theta,\psi$.
	\FOR{episode $j=1$ \TO $\infty$} \STATE{Initialize $k = 0$. Observe initial wealth $w_0$, initial stock price $s_0$, initial instantaneous variance $g_0$. Store $w_{t_k} \leftarrow  w_0$, $s_{t_k} \leftarrow  s_0$, $g_{t_k} \leftarrow  g_0$.
		\WHILE{$k < K$} \STATE{
			Generate action $a_{t_k}\sim \bm{\bm\pi}^{\psi}(\cdot|t_k,g_{t_k})$.
			
			Apply $a_{t_k}$ to market simulator $(s',g') = Market_{\Delta t}(t_k, s_{t_k}, g_{t_k})$, and observe new state $s',g'$. Store $s_{t_{k+1}} \leftarrow s'$, $g_{t_{k+1}} \leftarrow g'$.
			
			Compute  current wealth $w_{t_{k+1}} = w_{t_k} + w_{t_k}a_{t_k}\frac{s_{t_{k+1}}}{s_{t_k}} + w_{t_k}(1 - a_{t_k})r\Delta t $.
			
			Compute and store
			\[ \begin{aligned}
				\delta_{t_k} = & \frac{\hat V^{\psi}(t_{k+1},w_{t_{k+1}},g_{t_{k+1}}) - \hat V^{\psi}(t_{k},w_{t_{k}},g_{t_k})}{(1-\gamma) \hat V^{\psi}(t_{k},w_{t_{k}},g_{t_k}) + 1}  .
			\end{aligned} \]
			
			Update $k \leftarrow k + 1$
		}
		\ENDWHILE

		Update $\theta$ and $\psi$ by
		\[ \psi \leftarrow \psi + \ell(j)l_{\psi} \sum_{k=0}^{K-1} \delta_{t_k} \frac{\partial \hat V^{\psi}}{\partial \psi}(t_k,w_{t_k},g_{t_k}).  \]
		\[ \theta \leftarrow \theta + \ell(j)l_{\theta} \sum_{k=0}^{K-1}\delta_{t_k} \frac{\partial }{\partial \theta}\log\hat{\bm\pi}^{\theta}(a_{t_k}|t_k,g_{t_k}) .\]
	}
	\ENDFOR
\end{algorithmic}
\label{algo:offline incremental}
\end{algorithm}

\newpage
\section{Additional Numerical Results}
\label{appendix:additional numerical results}
In the main paper,  we generate a training dataset with a length of 20 years, and each time, we sample a subsequence with a length of 1 year as one episode for training. This is to capture the practical situation in which financial data are always limited. However, in a simulation study, we can generate as much data as we desire.  Here, we report the results of such a ``thought experiment" when data are unlimited. Specifically,  in each episode, we generate {\it independent} one-year data from the given dynamics for training. Figure \ref{fig:simulated payoff inf data no gbm} illustrates the learning curves of the two RL methods, where average utilities are computed on an independent, fixed test set with 10000 wealth trajectories. Both curves, based on specific forms and neural networks, converge to the omniscient optimal utility after about 3000 independent episodes.

\begin{figure}[H]
\centering
\includegraphics[width=0.5\textwidth]{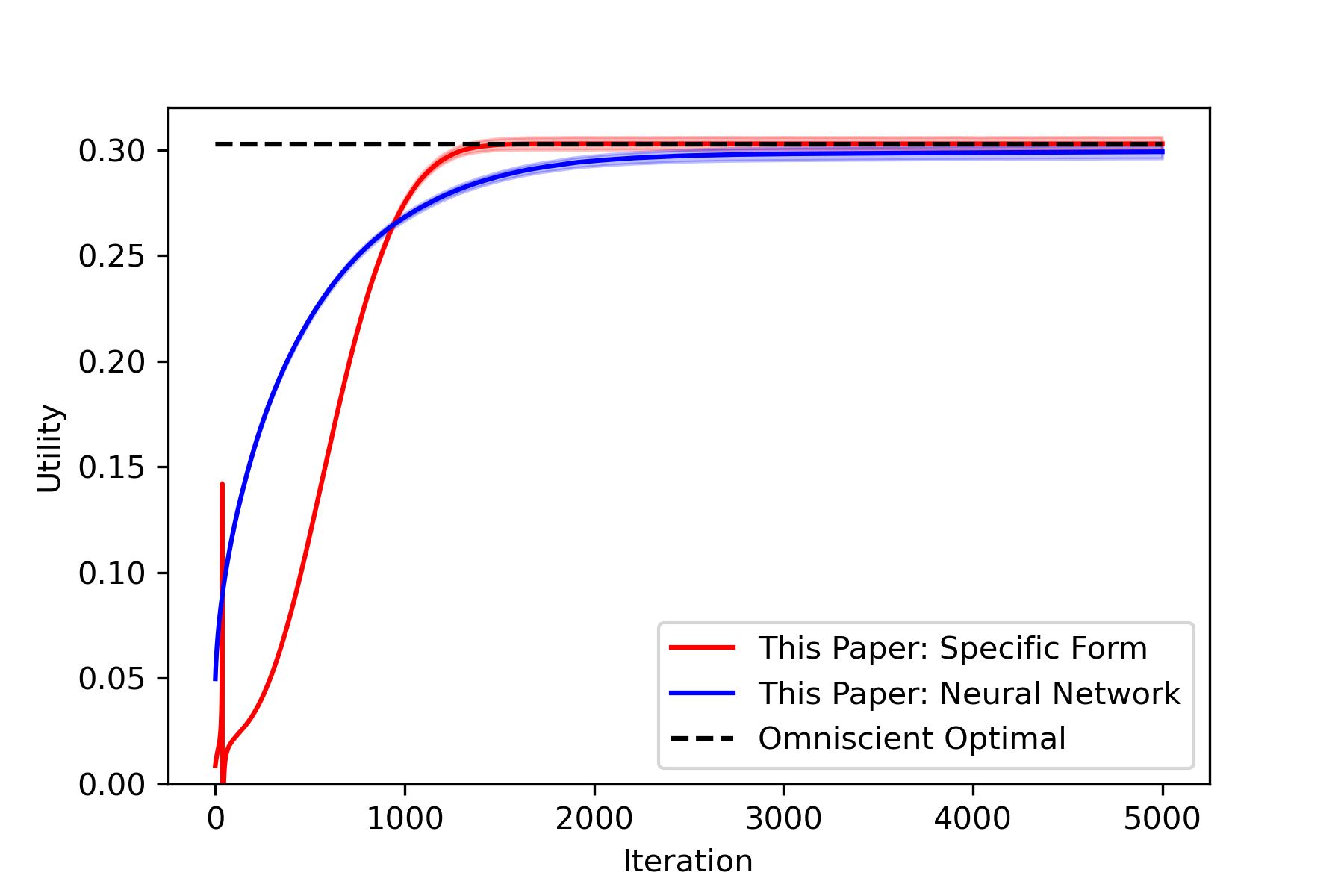}
\caption{\textbf{Average utility  of RL algorithms on the test set as functions of the number of
		training episodes.}  In each  episode,  independent one-year data are generated from the model dynamics for training. The width of the shaded area is twice the standard deviation.}
\label{fig:simulated payoff inf data no gbm}
\end{figure}

\section{Proof of Statements}

\subsection{Proof of Theorem \ref{thm:hjb solution}}
We first verify that the function $V^{(\lambda)}$ given in \eqref{eq:optimal value function} solves the HJB equation \eqref{eq:hjb 0}. Note that $V^{(\lambda)}_{ww} = -\gamma w^{1-\gamma} \exp\{ \varphi(t,x) - \lambda(1-\gamma)(T-t)/2 \} < 0$, hence the ``sup'' in \eqref{eq:hjb 0} is achieved at
\[ \bm u^*(t,x,w) = \frac{(\mu(t,x) - r)V^{(\lambda)}_{w} + \rho\nu(t,x)\sigma(t,x)  V^{(\lambda)}_{wx}}{-\sigma^2(t,x) w V^{(\lambda)}_{ww}} = \frac{\mu(t,x) - r}{\gamma \sigma^2(t,x)} +\frac{\rho\nu(t,x)}{\gamma\sigma(t,x)}\varphi_x(t,x) . \]
It is then straightforward to verify that $V^{(\lambda)}$ satisfies \eqref{eq:hjb 0}.

The rest of the results can be proved following a standard verification approach. We include the  proof for reader's convenience.

We first show that for any admissible policy $\bm\pi^{(\lambda)}$, the associated value function $J^{(\bm\pi^{(\lambda)})}$ defined in \eqref{eq:objective new} is smaller than $V^{(\lambda)}$. Let $(W^{\bm\pi^{(\lambda)}},X)$ be the wealth-factor process under $\bm\pi^{(\lambda)}$. Apply It\^o's lemma to $V^{(\lambda)}(t,W_t^{\bm\pi^{(\lambda)}},X_t)$ to obtain
\[ \begin{aligned}
& V^{(\lambda)}(T,W_T^{\bm\pi^{(\lambda)}},X_T) - V^{(\lambda)}(t,W_t^{\bm\pi^{(\lambda)}},X_t) \\
= & \int_t^T \dd s  \Bigg\{ \frac{\partial V^{(\lambda)} }{\partial t} + \left[r + (\mu(s,X_s) - r) u_s \right] W_s^{\bm\pi^{(\lambda)}} V^{(\lambda)}_{w} + \frac{1}{2}\sigma^2(s,X_s) \left( u_s^2 + \frac{\lambda}{\gamma \sigma^2(s,X_s)} \right) W_s^{\bm\pi^{(\lambda)} 2} V^{(\lambda)}_{ww}  \\
& \quad \quad \quad \quad + m(s,X_s)V^{(\lambda)}_{x} + \frac{1}{2}\nu^2(s,X_s) V^{(\lambda)}_{xx} + \rho\nu(s,X_s)\sigma(s,X_s) u_s W_s^{\bm\pi^{(\lambda)}} V^{(\lambda)}_{wx} \Bigg\} \\
& + \int_t^T \bigg\{ \sigma(s,X_s) u_s W_s^{\bm\pi^{(\lambda)}} V^{(\lambda)}_{w} \dd B_t  +  \sqrt{\frac{\lambda}{\gamma}} W_s^{\bm\pi^{(\lambda)}} V^{(\lambda)}_{w} \dd \bar{B}_t + \nu(s,X_s)V^{(\lambda)}_{x}[\rho\dd B_t + \sqrt{1-\rho^2}\dd \tilde B_t] \bigg\} ,
\end{aligned} \]
where  $u_s=\operatorname{Mean}(\bm\pi^{(\lambda)}(\cdot|s,W_s^{\bm\pi^{(\lambda)}},X_s))$.
Define a sequence of increasing stopping times $\tau_n = \inf\{ s\geq t: |X_s| \geq n \text{ or } \left( W_s^{\bm\pi^{(\lambda)}} \right)^{1-\gamma} \geq n \}$. Replacing $T$ by $T\wedge \tau_n$ in the above formula and taking conditional expectation on both sides, we obtain
\[\begin{aligned}
& \E\left[ V^{(\lambda)}(T\wedge \tau_n,W_{T\wedge \tau_n}^{\bm\pi^{(\lambda)}},X_{T\wedge \tau_n}) \mid W^{\bm\pi^{(\lambda)}}_t = w, X_t = x  \right] - V^{(\lambda)}(t,w,x)  \\
=& \E\Bigg[ \int_t^T \dd s  \Bigg\{ \frac{\partial V^{(\lambda)} }{\partial t} + \left[r + (\mu(s,X_s) - r) u_s \right] W_s^{\bm\pi^{(\lambda)}} V^{(\lambda)}_{w} + \frac{1}{2}\sigma^2(s,X_s) \left( u_s^2 + \frac{\lambda}{\gamma \sigma^2(s,X_s)} \right) W_s^{\bm\pi^{(\lambda)} 2} V^{(\lambda)}_{ww} \\
& \quad \quad \quad \quad + m(s,X_s)V^{(\lambda)}_{x} + \frac{1}{2}\nu^2(s,X_s) V^{(\lambda)}_{xx} + \rho\nu(s,X_s)\sigma(s,X_s) u_s W_s^{\bm\pi^{(\lambda)}} V^{(\lambda)}_{wx} \Bigg\} \mid W^{\bm\pi^{(\lambda)}}_t = w, X_t = x  \Bigg] \\
\leq & 0 ,
\end{aligned}  \]
where the last inequality is due to the fact that $V^{(\lambda)}$ solves the HJB equation \eqref{eq:hjb 0}, and the $\dd B_t, \dd \tilde B_t$ terms vanish because they are martingales. Moreover,
\[ \begin{aligned}
& \E\left[ V^{(\lambda)}(T\wedge \tau_n,W_{T\wedge \tau_n}^{\bm\pi^{(\lambda)}},X_{T\wedge \tau_n}) \mid W^{\bm\pi^{(\lambda)}}_t = w, X_t = x  \right] 	\\
= & -\frac{1}{1-\gamma} + \E\left[ \frac{\left( W^{\bm\pi^{(\lambda)}}_T  \right)^{1-\gamma}}{1-\gamma} \one_{\{ \tau_n  > T \}} \mid W^{\bm\pi^{(\lambda)}}_t = w, X_t = x  \right] \\
& + \E\left[ \one_{\{ \tau_n  \leq  T \}}  \frac{\left( W^{\bm\pi^{(\lambda)}}_{\tau_n}  \right)^{1-\gamma} \exp\{ \varphi(\tau_n, X_{\tau_n}) - \lambda (1-\gamma) (T-\tau_n)/2 \} }{1-\gamma} \mid W^{\bm\pi^{(\lambda)}}_t = w, X_t = x  \right] \\
= :& -\frac{1}{1-\gamma} + \frac{1}{1-\gamma} I_1 + \frac{1}{1-\gamma}I_2 .
\end{aligned} \]

By the monotone convergence theorem, we have
\[ \lim_{n\to\infty} I_1 =  \E\left[ \left( W^{\bm\pi^{(\lambda)}}_T  \right)^{1-\gamma}\mid W^{\bm\pi^{(\lambda)}}_t = w, X_t = x  \right] . \]
Thus, we have proved
\begin{equation}\label{midineq} V^{(\lambda)}(t,w,x) \geq  \E\left[ U\left( W^{\bm\pi^{(\lambda)}}_T  \right)\mid W^{\bm\pi^{(\lambda)}}_t = w, X_t = x  \right]  + \limsup_{n\to\infty} \frac{1}{1-\gamma}I_2 . \end{equation}

It suffices to show that $\lim_{n\to\infty} I_2 = 0 $. By H\"older's inequality, we have
\[ I_2 \leq n e^{\lambda T |1-\gamma|} \left( \p(\tau_n\leq T) \right)^{1/q} \left( \E\left[  \exp\{ p \varphi(T\wedge \tau_n, X_{T\wedge \tau_n}) \} \mid W^{\bm\pi^{(\lambda)}}_t = w, X_t = x \right] \right)^{1/p} ,  \]
for any $p,q>1$ satisfying $1/p + 1/q = 1$. By the regularity of the function $\varphi$, we have
\[ \limsup_{n\to\infty}\E\left[  \exp\{ p \varphi(T\wedge \tau_n, X_{T\wedge \tau_n}) \} \mid W^{\bm\pi^{(\lambda)}}_t = w, X_t = x \right] < \infty. \]
Moreover, standard growth conditions of SDEs yield
\[ \p(\tau_n\leq T) \leq C n^{-L} \]
where $L$ can be arbitrarily small. This implies that $\lim_{n\to\infty} I_2 = 0$.

Next, when the policy \eqref{eq:optimal exploratory policy via u} is taken, then the inequality  \eqref{midineq} becomes an equality, because \eqref{eq:optimal exploratory policy via u} achieves the supremum in the HJB equation \eqref{eq:hjb 0}  and the policy is admissible based on Definition \ref{def:admissible feedback}. This establishes the optimality of \eqref{eq:optimal exploratory policy via u}. Finally, the above analysis applies to the case when $\lambda=0$ noting that $\bm u^*$ is independent of $\lambda$. This proves the last statement and completes the proof.

\subsection{Proof of Corollary \ref{coro:erwl}}
It follows from the form of the optimal value function \eqref{eq:optimal value function} that
\[ V^{(\lambda)}(t,w,x) = V^{(0)}(t, \exp\{\frac{-\lambda (T-t)}{2}\} w,x) .\]
The desired result follows from Definition \ref{def:erwl}.

\subsection{Proof of Theorem \ref{lemma:value function one policy}}
\label{appendix:lemma1}
\begin{enumerate}[(i)]
\item Given the policy $\bm\pi^{(\lambda)}$, it follows from the Feynman-Kac formula that the value function $J^{(\bm\pi^{(\lambda)})}$ satisfies the linear PDE
\[ \begin{aligned}
	& \frac{\partial J^{(\bm\pi^{(\lambda)})} }{\partial t} +  \Big( r + \big(\mu(t,x) - r\big) \bm u(t,x) \Big)w J^{(\bm\pi^{(\lambda)})}_w + \frac{1}{2}\sigma^2(t,x)\Big( \bm u^2(t,x) +  \frac{\lambda}{\gamma \sigma(t,x)^2}\Big) w^2J^{(\bm\pi^{(\lambda)})}_{ww} \\
	& + m(t,x) J^{(\bm\pi^{(\lambda)})}_x + \frac{1}{2}\nu^2(t,x)J^{(\bm\pi^{(\lambda)})}_{xx} + \rho\nu(t,x)\sigma(t,x)\bm u(t,x) w J^{(\bm\pi^{(\lambda)})}_{wx}  = 0
\end{aligned} \]
with $J^{(\bm\pi^{(\lambda)})}(T,w,x)=U(w)$.
A direct calculation verifies that the function $J^{(\bm\pi^{(\lambda)})}$ specified in the statement satisfies the above PDE. The desired result then follows from the uniqueness of the solution to the linear PDE.

%\begin{equation} \label{Jtwx} J^{(\bm\pi^{(\lambda)})}(t, w, x) =  \frac{w^{1-\gamma} \exp\{ \bar\varphi(t,x) - \lambda (1-\gamma)(T-t)/2 \} - 1}{1-\gamma} , \end{equation}
%where $\bar\varphi$ satisfies the PDE
%\begin{equation} \begin{aligned}\label{PDEbar}
%\frac{\partial \bar\varphi}{\partial t} & + (1-\gamma)r  + m(t,x)\bar\varphi_x + \frac{1}{2}\nu^2(t,x)\left(\bar\varphi_{xx} + \left( \bar\varphi_x \right)^2 \right) \\
%& + (1-\gamma)\left[ (\mu(t,x) - r) \bm u(t,x) - \frac{\gamma}{2}\sigma(t,x)^2 \bm u(t,x)^2  + \rho \sigma(t,x)\nu(t,x) \bm u(t,x) \bar\varphi_x \right]  = 0,
%\end{aligned} \end{equation}
%\tilde{\bm u}

\item %Note that the PDE satisfied by $\varphi^{(\bm\pi^{(\lambda)})}$ does not depend on $\lambda$. Thus, we may write
%\[ J^{( \bm\pi^{(\lambda)}) }(t,w,x) =  \frac{w^{1-\gamma}\exp\{ \varphi^{(\bm\pi^{(0)})} - \lambda(1-\gamma)(T-t)/2 \}   - 1 }{1-\gamma} . \]
By (i), $J^{( \tilde{\bm\pi}^{(\lambda)}) }$ has the same representation \eqref{Jtwx} with $\bar\varphi$ replaced by $\tilde\varphi$ while the latter satisfies the PDE \eqref{PDEbar} with ${\bm u}$ replaced by $\tilde{\bm u}$.
Therefore, it suffices to show that $ \tilde\varphi \geq \bar\varphi$ when $0<\gamma<1$, and $ \tilde\varphi \leq \bar\varphi$ when $\gamma > 1$.

Consider the transformation: $\phi(t,w,x) = e^{\bar\varphi(t,w,x)}$. Then, $\phi$ satisfies the PDE
\[ \begin{aligned}
	& \frac{\partial \phi}{\partial t} + m(t,x) \phi_x + \frac{1}{2}\nu^2(t,x)\phi_{xx} \\
	& + (1-\gamma) \left[ r + (\mu(t,x) - r)  \bm u(t,x) \phi - \frac{\gamma}{2}\sigma^2(t,x)\bm u^2(t,x) \phi +\rho\nu(t,x)\sigma(t,x)\bm u(t,x) \phi_x \right] = 0 .
\end{aligned}\]
Similarly, $\tilde\phi(t,w,x) = e^{\tilde\varphi(t,w,x)}$ satisfies
\[ \begin{aligned}
	& \frac{\partial \tilde\phi}{\partial t} + m(t,x) \tilde\phi_x + \frac{1}{2}\nu^2(t,x)\tilde\phi_{xx} \\
	& + (1-\gamma) \left[ r + (\mu(t,x) - r)  \tilde{\bm u}(t,x) \tilde\phi - \frac{\gamma}{2}\sigma^2(t,x)\tilde{\bm u}^2(t,x) \tilde\phi +\rho\nu(t,x)\sigma(t,x)\tilde{\bm u}(t,x) \tilde\phi_x \right] = 0 ,
\end{aligned}\]
with $\tilde{\bm u}(t,x) = \frac{( \mu(t,x) - r) \bm u(t,x) \phi(t,x) + \rho \nu(t,x)\sigma(t,x) \bm u(t,x)\phi_{x}(t,x)}{\gamma\sigma^2(t,x) \phi(t,x)}$.
Note that
\[\begin{aligned}
	& (\mu(t,x) - r)  \bm u(t,x) \phi - \frac{\gamma}{2}\sigma^2(t,x)\bm u^2(t,x) \phi +\rho\nu(t,x)\sigma(t,x)\bm u(t,x) \phi_x \\
	\leq & (\mu(t,x) - r)  \tilde{\bm u}(t,x) \phi - \frac{\gamma}{2}\sigma^2(t,x)\tilde{\bm u}^2(t,x) \phi +\rho\nu(t,x)\sigma(t,x)\tilde{\bm u}(t,x) \phi_x .
\end{aligned}   \]
Therefore, when $0<\gamma<1$, we have
\[ \begin{aligned}
	& \frac{\partial \phi}{\partial t} + m(t,x) \phi_x + \frac{1}{2}\nu^2(t,x)\phi_{xx} \\
	& + (1-\gamma) \left[ r + (\mu(t,x) - r)  \tilde{\bm u}(t,x) \phi - \frac{\gamma}{2}\sigma^2(t,x)\tilde{\bm u}^2(t,x) \phi +\rho\nu(t,x)\sigma(t,x)\tilde{\bm u}(t,x) \phi_x\right] \geq 0 .
\end{aligned} \]

By the comparison principle of PDEs, we have $\tilde\phi \geq \phi$ when $0<\gamma<1$. The case for $\gamma > 1$ can be proved in parallel. This completes the proof.
\end{enumerate}

\subsection{Proof of Theorem \ref{thm:pg theory}} \label{appendix:pg theory}
(i) The equation in the statement   is essentially the martingale orthogonality condition for policy evaluation developed in \citet{jia2021policy}.  Following the same argument  as in the proof of Proposition 4 therein, we obtain
\[ \hat V(t_0,w_0,x_0) = \E\left[ U(W_T^{a^{\bm\pi^{(\lambda)}}}) \big| W_{t_0}^{a^{\bm\pi^{(\lambda)}}} = w_0, X_{t_0} = x_0  \right], \]
which, by definition, coincides with the value function $J^{(\bm\pi^{(\lambda)})}$.

(ii)
Denote $\mu_t = \mu(t,X_t)$, $\sigma_t = \sigma(t,X_t)$, $\eta_t=\eta(t, W_t^{a^{\hat{\bm\pi}^{(\lambda)}}}, X_t)$, and
\[ J_t = J^{(\bm\pi^{(\lambda)})}(t, W_t^{a^{\hat{\bm\pi}^{(\lambda)}}}, X_t)  = \frac{\left(  W_t^{a^{\hat{\bm\pi}^{(\lambda)}}} \right)^{1-\gamma} \exp\{ \varphi(t,X_t)  - \lambda(1-\gamma)(T-t)/2 \}  - 1}{1-\gamma},\ \varphi_t = \varphi(t,X_t) . \]
Apply It\^o's lemma to $J_t$ to obtain
\[ \begin{aligned}
&	\E\left[ \int_{t_0}^T \eta_t \left( a_t^{\hat{\bm\pi}^{(\lambda)}} - \hat{\bm u}(t,X_t)\right) \dd J^{(\bm\pi^{(\lambda)})}(t, W_t^{a^{\hat{\bm\pi}^{(\lambda)}}}, X_t)   \right] \\
= & \E\Bigg[ \int_{t_0}^T \eta_t \left( a_t^{\hat{\bm\pi}^{(\lambda)}} - \hat{\bm u}(t,X_t)\right) \left(  W_t^{a^{\hat{\bm\pi}^{(\lambda)}}} \right)^{1-\gamma} \exp\{ \varphi_t  - \lambda(1-\gamma)(T-t)/2 \}  \times \\
& \quad \bigg\{ \Big[ \left( r + (\mu_t-r) a_t^{\hat{\bm\pi}^{(\lambda)}} - \frac{\gamma}{2}\sigma^2_t (a_t^{\hat{\bm\pi}^{(\lambda)}})^2  \right)  \dd t + \cdots +\dd B_t + \sigma_t a_t^{\hat{\bm\pi}^{(\lambda)}} \dd \langle B ,\varphi\rangle_t  \Big] \Bigg.\\
&\quad \Bigg. + \frac{1}{1-\gamma}\left[ \dd \varphi_t + \frac{1}{2}\dd \langle \varphi \rangle_t + \frac{\lambda(1-\gamma)}{2}\dd t \right]  \bigg\}   \Bigg] \\
= & \E\Bigg[ \int_{t_0}^T \eta_t \left( a_t^{\hat{\bm\pi}^{(\lambda)}} - \hat{\bm u}(t,X_t)\right) \left(  W_t^{a^{\hat{\bm\pi}^{(\lambda)}}} \right)^{1-\gamma} \exp\{ \varphi_t  - \lambda(1-\gamma)(T-t)/2 \}  \times \\
& \quad  \bigg\{ \Big[ \left( (\mu_t-r) \left( a_t^{\hat{\bm\pi}^{(\lambda)}} - \hat{\bm u}(t,X_t)\right) - \gamma \sigma^2_t\hat{\bm u}(t,X_t) \left( a_t^{\hat{\bm\pi}^{(\lambda)}} - \hat{\bm u}(t,X_t)\right)  \right)  \dd t + \sigma_t \left( a_t^{\hat{\bm\pi}^{(\lambda)}} - \hat{\bm u}(t,X_t)\right) \dd \langle B ,\varphi\rangle_t  \Big]   \bigg\}   \Bigg] \\
= & \E\left[ \int_{t_0}^T \eta_t \frac{\lambda}{\gamma \sigma^2_t} \left(  W_t^{a^{\hat{\bm\pi}^{(\lambda)}}} \right)^{1-\gamma} \exp\{ \varphi_t  - \lambda(1-\gamma)(T-t)/2 \}    \left[ (\mu_t-r) \dd t - \gamma \sigma^2_t \hat{\bm u}(t,X_t)   \dd t + \sigma_t \dd \langle B ,\varphi\rangle_t \right]     \right] ,
\end{aligned} \]
where $ \langle B ,\varphi\rangle$ is the covariational process between $B_t$ and $\varphi_t$ and $ \langle \varphi\rangle$ is the quadratic variation process of $\varphi_t$. Hence
\[ \dd \langle B ,\varphi\rangle_t  = \rho \nu(t,X_t) \varphi_x(t,X_t) \dd t . \]

Since the above expectation equals zero for any test process $\eta$, the integrand is zero almost surely. Therefore, we have
\[  \mu_t-r - \gamma \sigma^2_t \hat{\bm u}(t,X_t) + \rho\sigma_t \nu(t,X_t) \varphi_x(t,X_t)  = 0  \]
or
\[ \hat{\bm u}(t,X_t) = \frac{  \mu(t,X_t) - r  + \rho\nu(t,X_t)\sigma(t,X_t) \varphi_x(t,X_t)  }{\gamma \sigma^2(t,X_t)} = \tilde{\bm u}(t,X_t)  \]
almost surely and almost all $t\in[t_0,T]$. Because both $\hat{\bm u}$ and $\tilde{\bm u}$ are continuous function, we conclude $\hat{\bm u}(t_0,x_0)=\tilde{\bm u}(t_0,x_0)$. This completes the proof because $(t_0,x_0)$ are arbitrary.

\subsection{Proof of Proposition \ref{prop:gbm discrete time}}
It follows from  the wealth equation \eqref{classical_wealth} that
\[ \log\left(W_{t_{k+1}}/W_{t_k}\right) = [r + (\mu-r)a_{t_k} - \frac{1}{2}\sigma^2 a_{t_k}^2]\Delta t + a_{t_k}\sigma (B_{t_{k+1}} - B_{t_k}). \]
%For simplicity, denote
%\[ L_k = \left(W_{t_{k+1}}/W_{t_k}\right)^{1-\gamma}\exp\left\{ \left[-\psi + \lambda(1-\gamma)/2 \right]\Delta t \right\} . \]
Hence
\[\begin{aligned}
& \E\left[ \left(W_{t_{k+1}}/W_{t_k}\right)^{1-\gamma} \big| a_{t_k}, W_{t_k} \right] \\
= & \exp\left\{(1-\gamma)[ r + (\mu-r)a_{t_k} - \frac{1}{2}\sigma^2 a_{t_k}^2]\Delta t +\frac{1}{2}(1-\gamma)^2 a_{t_k}^2\sigma^2\Delta t  \right\} \\
= & \exp\left\{ (1-\gamma)\left[r + (\mu-r)\theta - \frac{\gamma}{2}\sigma^2\theta^2 + (\mu-r-\theta\gamma\sigma^2) (a_{t_k} - \theta) - \frac{\gamma}{2}\sigma^2(a_{t_k} - \theta)^2 \right]\Delta t \right\}.
\end{aligned}   \]
Denote
\[\begin{aligned}
& A_1 = (1-\gamma)\left[r + (\mu-r)\theta - \frac{\gamma}{2}\sigma^2\theta^2 + \frac{1}{2}\lambda \right] - \psi,\ A_2 = (1-\gamma)(\mu-r-\theta\gamma\sigma^2)\sqrt{\frac{\lambda}{\gamma\sigma^2}} \\
& A_3 = (\gamma-1)\lambda/2, \ A_4 = 2(1-\gamma)\left[r + (\mu-r)\theta - \frac{2\gamma-1}{2}\sigma^2\theta^2 + \frac{1}{2}\lambda \right] - 2\psi \\
& A_5 = 2(1-\gamma)(\mu-r-\frac{2\gamma-1}{2}\theta\sigma^2)\sqrt{\frac{\lambda}{\gamma\sigma^2}}, \ A_6 = (\gamma-1)(2\gamma-1)\lambda/\gamma.
\end{aligned}    \]

Let $L_k = \left(W_{t_{k+1}}/W_{t_k}\right)^{1-\gamma}\exp\left\{ \left[-\psi + \lambda(1-\gamma)/2 \right]\Delta t \right\}$. Then
\[\begin{aligned}
& \E\left[ L_k-1\right] \\
= &\E\left[ \E\left[ \left(W_{t_{k+1}}/W_{t_k}\right)^{1-\gamma}\exp\left\{ \left[-\psi + \lambda(1-\gamma)/2 \right]\Delta t \right\}     - 1 \big| a_{t_k},W_{t_k} \right] \right] \\
= & \E\bigg[ \exp\left\{ (1-\gamma)\left[ (\mu-r-\theta\gamma\sigma^2) (a_{t_k} - \theta) - \frac{\gamma}{2}\sigma^2(a_{t_k} - \theta)^2 \right]\Delta t \right\} e^{A_1 \Delta t} - 1 \bigg] \\
= & e^{A_1\Delta t} \frac{\exp\left\{ \frac{A_2^2(\Delta t)^2}{2[1 - A_3 \Delta t]} \right\}}{\sqrt{1 - \lambda(\gamma - 1)\Delta t}}   - 1
\end{aligned}   \]
and, therefore,
\[\begin{aligned}
& \E\left[ (a_{t_k} - \theta) (L_k-1)\right] \\
= &\E\left[ (a_{t_k} - \theta) \E\left[ \left(W_{t_{k+1}}/W_{t_k}\right)^{1-\gamma}\exp\left\{ \left[-\psi + \lambda(1-\gamma)/2 \right]\Delta t \right\}     - 1 \big| a_{t_k},W_{t_k} \right] \right] \\
= & \E\bigg[ (a_{t_k} - \theta) \left( \exp\left\{ (1-\gamma)\left[ (\mu-r-\theta\gamma\sigma^2) (a_{t_k} - \theta) - \frac{\gamma}{2}\sigma^2(a_{t_k} - \theta)^2 \right]\Delta t \right\}e^{A_1\Delta t} - 1 \right) \bigg] \\
= & \sqrt{\frac{\lambda}{\gamma\sigma^2}} e^{A_1\Delta t} A_2\Delta t \frac{\exp\left\{ \frac{A_2^2(\Delta t)^2}{2[1 - A_3 \Delta t]} \right\}}{\left(  1 - A_3\Delta t\right)^{3/2}}  .
\end{aligned}   \]
Similarly, we obtain
\[ \begin{aligned}
& \E\left[ (a_{t_k} - \theta)^2 L_k \right] = \frac{\lambda}{\gamma\sigma^2}e^{A_1\Delta t} \left[ 1 + A_2^2(\Delta t)^2 - 2 A_3\Delta t \right]  \frac{\exp\left\{ \frac{A_2^2(\Delta t)^2}{2[1 - A_3 \Delta t]} \right\}}{\left(  1 - A_3\Delta t\right)^{5/2}} .
\end{aligned} \]
\[ \begin{aligned}
& \E\left[ (a_{t_k} - \theta)^2 L_k^2 \right]= \frac{\lambda}{\gamma\sigma^2}e^{A_4\Delta t} \left[1 + A_5^2(\Delta t)^2 - 2A_6\Delta t  \right]  \frac{\exp\left\{ \frac{A_5^2(\Delta t)^2}{2[1 - A_6 \Delta t]} \right\}}{\left(  1 - A_6\Delta t\right)^{5/2}} .
\end{aligned} \]

Now we can compute
\[\begin{aligned}
& \E\left[ \widehat{e(\psi,\theta)} \right] \\
= & \sum_{k=0}^{K-1} \E\left[ \frac{\gamma \sigma^2 (a_{t_k} - \theta)}{\lambda (1-\gamma)} \left[ \left( \frac{W_{t_{k+1}}}{W_{t_{k}}} \right)^{1-\gamma} \exp\left\{ \left[-\psi + \lambda(1-\gamma)/2 \right]\Delta t \right\}     - 1\right] \right] \\
= &\frac{\gamma\sigma^2 K}{\lambda(1-\gamma)}\E\left[ (a_{t_k} - \theta) \left( L_k - 1 \right) \right] \\
= & \frac{\gamma\sigma^2 K}{\lambda(1-\gamma)} \sqrt{\frac{\lambda}{\gamma\sigma^2}}  e^{A_1\Delta t} A_2\Delta t \frac{\exp\left\{ \frac{A_2^2(\Delta t)^2}{2[1 - A_3 \Delta t]} \right\}}{\left(  1 - A_3\Delta t\right)^{3/2}}  \\
= & T(\mu-r-\theta\gamma\sigma^2)  \frac{\exp\left\{A_1 \Delta t + \frac{A_2^2(\Delta t)^2}{2[1 - A_3 \Delta t]} \right\}}{\left(  1 - A_3\Delta t\right)^{3/2}} .
\end{aligned}  \]

Under the conditions provided on $\Delta t$, there exists a constant $C$ that only depends on $\mu,r,\sigma,T$ such that $|A_1 \Delta t|\leq C(\mu,r,\sigma,\gamma,T), A_2^2(\Delta t)^2\leq C(\mu,r,\sigma,\gamma, T)$, and $1 - A_3\Delta t \in (\frac{3}{4}, \frac{5}{4})$. Thus,
%\[\frac{\exp\left\{A_1 \Delta t + \frac{A_2^2(\Delta t)^2}{2[1 - A_3 \Delta t]} \right\}}{\left(  1 - A_3\Delta t\right)^{3/2}}   - 1 \geq \frac{1}{\left( 1 - A_3\Delta t\right)^{3/2}}\left( A_1 \Delta t + \frac{A_2^2(\Delta t)^2}{2[1 - A_3 \Delta t]} \right) \geq -C |A_1|\Delta t  ,\]
%\[ \frac{\exp\left\{A_1 \Delta t + \frac{A_2^2(\Delta t)^2}{2[1 - A_3 \Delta t]} \right\}}{\left(  1 - A_3\Delta t\right)^{3/2}}   - 1 \leq \left(1 + C |A_1| \Delta t + C A_2^2(\Delta t)^2  \right)(1 + C |A_3|\Delta t) - 1 \leq C(|A_1| + |A_3|)\Delta t .   \]

Denote $f(\Delta t) = \exp\left\{A_1 \Delta t + \frac{A_2^2(\Delta t)^2}{2[1 - A_3 \Delta t]} \right\}$, $g(\Delta t) = \left(  1 - A_3\Delta t\right)^{-3/2}$. Notice that $f'(\Delta t) = \exp\left\{A_1 \Delta t + \frac{A_2^2(\Delta t)^2}{2[1 - A_3 \Delta t]} \right\} \left( A_1 + \frac{A_3A_2^2(\Delta t)^2}{2	(1-A_3 \Delta t)^2} + \frac{A_2^2\Delta t}{1-A_3\Delta t} \right)$, and $g'(\Delta t) = \frac{3}{2}A_3 \left(  1 - A_3\Delta t\right)^{-5/2}$. Then,
\[\begin{aligned}
& |f(\Delta t) g(\Delta t) - f(0)g(0)| \leq |f(\Delta t) - f(0)| g(\Delta t) + f(0)| g(\Delta t) - g(0) | \\
\leq & (\frac{3}{4})^{-3/2} \sup_{s: |A_1| s\leq C, A_2^2 s^2 \leq C, 1-A_3 s\in (\frac{3}{4}, \frac{5}{4})} |f'(s)| \Delta t + \sup_{s: |A_1| s\leq C, A_2^2 s^2 \leq C, 1-A_3 s\in (\frac{3}{4}, \frac{5}{4}) } |g'(s)| \Delta t \\
\leq & C_1 \left(  |A_1| + |A_2| + |A_3| \right) \Delta t .
\end{aligned}
\]

Therefore,
\[\begin{aligned}
& \left| \E\left[ \widehat{e(\psi,\theta)} \right] - T(\mu-r-\theta\gamma\sigma^2) \right| \\
\leq & T(\mu-r-\theta\gamma\sigma^2) \left| \frac{\exp\left\{A_1 \Delta t + \frac{A_2^2(\Delta t)^2}{2[1 - A_3 \Delta t]} \right\}}{\left(  1 - A_3\Delta t\right)^{3/2}}   - 1 \right| \\
\leq & C (1 + |\theta^2| +|\psi| + \lambda)  \Delta t .
\end{aligned}  \]

On the other hand,
\[ \begin{aligned}
& \operatorname{Var}\left[ \widehat{e(\psi,\theta)} \right] \\
= & \sum_{k=0}^{K-1} \operatorname{Var}\left[ \frac{\gamma \sigma^2 (a_{t_k} - \theta)}{\lambda (1-\gamma)} \left(L_k     - 1\right) \right] = \left( \frac{\gamma\sigma^2}{\lambda(1-\gamma)} \right)^2 K \operatorname{Var}\left[ (a_{t_k} - \theta) \left( L_k  - 1\right)  \right] \\
\leq & K\left( \frac{\gamma\sigma^2}{\lambda(1-\gamma)} \right)^2 \E\left[ (a_{t_k} - \theta)^2(L_k - 1)^2 \right] \\
= & K\left( \frac{\gamma\sigma^2}{\lambda(1-\gamma)} \right)^2 \frac{\lambda}{\gamma\sigma^2} \bigg[ e^{A_4\Delta t} \left[1 + A_5^2(\Delta t)^2 - 4A_6\Delta t  \right]  \frac{\exp\left\{ \frac{A_5^2(\Delta t)^2}{2[1 - A_6 \Delta t]} \right\}}{\left(  1 - A_6\Delta t\right)^{5/2}} \\
& - 2  e^{A_1\Delta t} \left[ 1 + A_2^2(\Delta t)^2 - 2 A_3\Delta t \right]  \frac{\exp\left\{ \frac{A_2^2(\Delta t)^2}{2[1 - A_3 \Delta t]} \right\}}{\left(  1 - A_3\Delta t\right)^{5/2}} + 1 \bigg].
\end{aligned} \]

Denote
\[ \begin{aligned}
h(\Delta t) = e^{A_4\Delta t} \left[1 + A_5^2(\Delta t)^2 - 4A_6\Delta t  \right]  \frac{\exp\left\{ \frac{A_5^2(\Delta t)^2}{2[1 - A_6 \Delta t]} \right\}}{\left(  1 - A_6\Delta t\right)^{5/2}} - 2  e^{A_1\Delta t} \left[ 1 + A_2^2(\Delta t)^2 - 2 A_3\Delta t \right]  \frac{\exp\left\{ \frac{A_2^2(\Delta t)^2}{2[1 - A_3 \Delta t]} \right\}}{\left(  1 - A_3\Delta t\right)^{5/2}} + 1 .
\end{aligned}  \]
Under the conditions provided on $\Delta t$, there exists a constant $C$ that only depends on $\mu,r,\sigma,T$ such that $(|A_1| + |A_4|)\Delta t\leq C(\mu,r,\sigma,\gamma,T), (A_2^2 + A_5^2)(\Delta t)^2\leq C(\mu,r,\sigma,\gamma, T)$, and $1 - A_3\Delta t, 1-A_6\Delta t \in (\frac{3}{4}, \frac{5}{4})$.
We have
\[\begin{aligned}
| h(\Delta t) |\leq & |h(0)| + |h'(0)|\Delta t + \frac{1}{2}\sup_{s:(|A_1| + |A_4|) s\leq C(\mu,r,\sigma,\gamma,T), (A_2^2 + A_5^2) s^2\leq C(\mu,r,\sigma,\gamma, T),1 - A_3 s, 1-A_6 s \in (\frac{3}{4}, \frac{5}{4}) }|h''(s)|(\Delta t)^2  \\
\leq & |A_4 - \frac{3}{2}A_6 - 2A_1-A_3| \Delta t + C(1 + |A_1|^2 + |A_2|^2 + \cdots + |A_6|^2)(\Delta t)^2.
\end{aligned} \]

Therefore,
\[\begin{aligned}
\operatorname{Var}\left[ \widehat{e(\psi,\theta)} \right] \leq & K \frac{\gamma\sigma^2}{\lambda (1-\gamma)^2}\left[  |A_4 - \frac{3}{2}A_6 - 2A_1-A_3| \Delta t + C(1+\psi^2+\theta^4+\lambda^2)(\Delta t)^2 \right] \\
\leq & C(1 + \frac{\theta^2}{\lambda}) + C\left(\frac{1+\psi^2+\theta^4}{\lambda}+\lambda\right) \Delta t .
\end{aligned}  \]
\subsection{Proof of Theorem \ref{thm:bs converge}}
We first prove a result  regarding equivalent relative wealth loss (ERWL) in the Black-Scholes market.
\begin{lemma}
\label{lemma:erwl bs}
In the Black-Scholes market, the equivalent relative wealth loss of a determinist policy $\bm u^{\theta} \equiv  \theta$ is
\[ \operatorname{ERWL}(\bm u^{\theta}) = 1 - \exp\{ -\frac{T\gamma \sigma^2}{2} (\theta - \theta^*)^2 \} \leq \frac{T\gamma \sigma^2}{2} (\theta - \theta^*)^2, \]
where $\theta^* = \frac{\mu-r}{\gamma\sigma^2}$ is the ground truth optimal allocation.
\end{lemma}
\proof{Proof of Lemma \ref{lemma:erwl bs}}
We show by direct calculation. Under the deterministic policy $\bm u^{\theta} = \theta$, it follows from  \eqref{classical_wealth} that the corresponding wealth process $W^\theta$ satisfies
\[ \frac{ \dd W_t^{\theta} }{W_t^{\theta}} = \theta \frac{\dd S_t}{S_t} + (1-\theta)r\dd t = [r + (\mu-r)\theta]\dd t + \sigma \theta \dd B_t . \]
Hence, $\log W_T^{\theta} \sim \mathcal N\left( \log w_0 + [r + (\mu-r)\theta - \frac{1}{2}\sigma^2\theta^2]T,\sigma^2\theta^2 T \right)$, leading to
\[\begin{aligned}
J(0,w_0) = & \frac{1}{1-\gamma} \E\left[ e^{(1-\gamma)\log  W_T^{\theta}} \right] - \frac{1}{1-\gamma} \\
= & \frac{w_0^{1-\gamma}}{1-\gamma} \exp\{ (1-\gamma) [r + (\mu-r)\theta - \frac{1}{2}\sigma^2\theta^2]T + \frac{(1-\gamma)^2}{2}\sigma^2\theta^2 T \} -\frac{1}{1-\gamma} \\
= &  \frac{w_0^{1-\gamma} \exp\{(1-\gamma) T [r + (\mu-r)\theta - \frac{\gamma}{2}\sigma^2\theta^2]   \}  - 1}{1-\gamma} \\
= & \frac{w_0^{1-\gamma} \exp\left\{-\frac{(1-\gamma)\gamma T\sigma^2}{2} (\theta - \theta^*)^2 + [r + \frac{(\mu-r)^2}{2\gamma\sigma^2}](1-\gamma) T  \right\}  - 1}{1-\gamma} \\
= & V^{(0)}\left( 0, w_0 e^{-\frac{\gamma T\sigma^2}{2} (\theta - \theta^*)^2} \right).
\end{aligned}  \]
Hence, by Definition \ref{def:erwl}, we have
\[ \operatorname{ERWL}(\bm u^{\theta}) = 1 - \exp\{ -\frac{T\gamma \sigma^2}{2} (\theta - \theta^*)^2 \} \leq \frac{T\gamma \sigma^2}{2} (\theta - \theta^*)^2 ,\]
where the inequality follows from the basic inequality $e^{z}\geq 1+z$ with $z = -\frac{T\gamma \sigma^2}{2} (\theta - \theta^*)^2$.
\endproof

The next lemma is about  a particular recursive relation.
\begin{lemma}
\label{lemma:recursive}
Suppose $\{e_n\}_{n\geq n_0}$ is a sequence of positive real numbers and $n_0\geq 4$ satisfying
\[ e_{n+1} \leq (1-\alpha_n) e_{n} + (C_1+C_2\log n) \alpha_n^2 ,\;\; \forall n\geq n_0,\]
where $\{\alpha_n\}_{n\geq 0}$ is a positive sequence satisfying $\alpha_n \leq \alpha_{n+1}(1 + A a_{n+1})$ for all $n\geq n_0$ and some $A\in (0,1)$. Let $C = \frac{1}{1-A}\sup_{n\geq n_0}\frac{C_1+C_2\log n}{\log (n-1)} \vee \frac{e_{n_0 + 1}}{\alpha_{n_0}}$. Then $e_{n+1}\leq C \alpha_n \log n$ $\forall n\geq n_0$.
\end{lemma}
\proof{Proof of Lemma \ref{lemma:recursive}}
We prove by induction. The conclusion holds for $n=n_0$ because $C \geq \frac{e_{n_0 + 1}}{\alpha_{n_0}}$. Assuming the conclusion holds for all $n \leq k$ with $k\geq n_0$, we examine the case with $n = k+1$. By the given recursive condition and the induction assumption, we have
\[\begin{aligned}
e_{k+1} \leq & (1-\alpha_k)e_k + (C_1 + C_2\log k) \alpha_k^2 \leq (1-\alpha_k) C \alpha_{k-1} \log(k-1) + (C_1+C_2\log k)\alpha_k^2 \\
= & C \alpha_{k} \log k \left( \alpha_{k-1} \frac{1 - \alpha_k}{\alpha_k}\log(k-1) + \alpha_k \frac{C_1+C_2\log k}{C \log k}  \right) \\
\leq & C \alpha_k\log k \left[ (1+A\alpha_k)(1-\alpha_k) \frac{\log(k-1)}{\log k}  + \alpha_k \frac{C_1+C_2\log k)}{C \log k} \right] \\
< & C\alpha_k\log k + C\alpha_k^2\log k \left( -A\alpha_k \frac{\log(k-1)}{\log k}- (1 -A)\frac{\log(k-1)}{\log k} + \frac{C_1+C_2\log k)}{C \log k} \right) \\
< & C\alpha_k \log k + \alpha_k^2  \left[ C_1+C_2\log k-(1-A)C \log(k-1)\right] < C \alpha_k \log k ,
\end{aligned}  \]
where the last inequality is because $C (1-A) > \sup_{n\geq n_0}\frac{C_1+C_2\log n}{\log (n-1)} \geq \frac{C_1+C_2\log k}{\log(k-1)}$. This proves the desired result.
\endproof

We are now ready to prove Theorem \ref{thm:bs converge}. By Lemma \ref{lemma:erwl bs}, we only need to focus on estimating $\E\left[ (\theta_n - \theta^*)^2 \right]$.

Recall that $\theta_n$ satisfies the recursion $\theta_{n+1} = \Pi_{K_{n+1}}\left( \theta_n + \ell_n \widehat{e_n(\psi,\theta)}\right)$, where $\widehat{e_n(\psi,\theta)}$ is specified in the statement of Theorem \ref{thm:bs converge}.
%a sample obtained as in Proposition \ref{prop:gbm discrete time} with $\theta_n,\psi_n,\Delta t_n$.
Hence, by Proposition \ref{prop:gbm discrete time}, we have
\[\begin{aligned}
& \left| \E\left[ \widehat{e_n(\psi,\theta)} \big| \theta_n,\phi_n \right] - h(\theta_n)  \right| \leq C (1+|\theta_n|+|\psi_n|+\lambda)\Delta t_n =:\beta_n \\
&  \operatorname{Var}\left[ \widehat{e_n(\psi,\theta)} - h(\theta_n) \big| \theta_n,\phi_n  \right] \leq C(1 + \frac{|\theta_n|^2}{\lambda}) + C(\frac{1 + |\psi_n|^2 + |\theta_n|^2}{\lambda} + \lambda) \Delta t_n =:\zeta_n
\end{aligned} ,\]
where $h(\theta) = T\gamma\sigma^2 (\theta^* - \theta)$, and $C$ is a constant that only depends on $\mu,r,\sigma,\gamma,T$.

Write $\widehat{e_n(\psi,\theta)} = h(\theta_n) + \xi_n$, where $\E\left[ \xi_n \big| \theta_n,\phi_n \right] \leq \beta_n$ and $\operatorname{Var}\left[ \xi_n \big| \theta_n,\phi_n  \right]\leq \zeta_n $. By the properties of the projection mapping, we have
\[\left| \theta_{n+1} - \theta^* \right|^2 \leq \left| \theta_n - \theta^* + \ell_n \left( h(\theta_n) + \xi_n \right)  \right|^2 . \]
Therefore,
\[ \begin{aligned}
& \E\left[ \left| \theta_{n+1} - \theta^* \right|^2 \big| \theta_n,\psi_n \right]\\
\leq & (1-\ell_n T\gamma\sigma^2)^2(\theta_n-\theta^*)^2 + 2\ell_n(\theta_n-\theta^*)\beta_n + 2 \ell_n^2 T\gamma\sigma^2(\theta^*-\theta_n)\beta_n +\ell_n^2(\beta_n^2 + \zeta_n) \\
\leq & (1-\ell_n T\gamma\sigma^2)^2(\theta_n-\theta^*)^2 + \ell_n(1 - \ell_n T\gamma\sigma^2)\left[ (\theta_n-\theta^*)^2 + 1\right]\beta_n  + \ell_n^2(\beta_n^2+\zeta_n) \\
= & (1-\ell_n T\gamma\sigma^2) (1 - \ell_n T\gamma\sigma^2 +  \ell_n \beta_n) (\theta_n-\theta^*)^2 +  \ell_n (1 - \ell_n T\gamma\sigma^2)\beta_n  + \ell_n^2(\beta_n^2+\zeta_n) .
\end{aligned}\]

By a property of projection, we know $|\theta_n| \leq c_n \leq \sqrt{\log n}$. Hence, for all $n\geq n_0$, we have
\[ \beta_n \leq C(\lambda + 1+M+c_n)\Delta t_n \leq C(\lambda+1+M+\log n)\Delta t_n \wedge 1, \]
and
\[ \zeta_n \leq C(1 + c_n^2 \lambda^{-1}) +  C  (\lambda + \frac{1+M^2 + c_n^2}{\lambda})\Delta t_n \leq 2C(1 +\lambda^{-1} \log n ) + C\lambda \Delta t_n . \]
Therefore,
\[ \begin{aligned}
& \E\left[ \left| \theta_{n+1} - \theta^* \right|^2 \big| \theta_n,\psi_n \right]\\
\leq & (1-\ell_n T\gamma\sigma^2) (1 - \ell_n T\gamma\sigma^2 +  \ell_n \beta_n) (\theta_n-\theta^*)^2 +  \ell_n(1 - \ell_n T\gamma\sigma^2)\beta_n  + \ell_n^2(\beta_n^2+\zeta_n) \\
\leq & (1-\ell_n T\gamma\sigma^2) (\theta_n-\theta^*)^2 + \ell_n C(\lambda+1+M+\log n)\Delta t_n + \ell_n^2\left( 2C +2C\lambda^{-1} \log n + C\lambda \Delta t_n  \right) .
\end{aligned} \]

Taking  expectation, we obtain a recursive relation for $\E\left[ (\theta_n - \theta^*)^2 \right]$:
\[\begin{aligned}
\E\left[ (\theta_{n+1} - \theta^*)^2 \right] \leq & (1-\ell_n T\gamma\sigma^2) \E\left[(\theta_n-\theta^*)^2\right] \\
& + \ell_n^2\left(  \frac{C(\lambda+1+M+\log n)\Delta t_n}{\ell_n} + 2C + 2C\lambda^{-1} \log n + C\lambda \Delta t_n  \right) \\
\leq & (1-\ell_n T\gamma\sigma^2) \E\left[(\theta_n-\theta^*)^2\right] + \ell_n^2 C_2 (1 + \lambda^{-1}\log n + \lambda),
\end{aligned} \]
where $C_2$ is a constant that only depends on $\mu,r,\sigma,\gamma,T$.

By the specification of $\ell_n$ in the condition, a direct calculation verifies that $\ell_n$ satisfies $ \ell_n\leq \ell_{n+1} \left(1 + \eta_2 \ell_{n+1}  \right) $, for all $n\geq n_0$. It follows now from  Lemma \ref{lemma:recursive} that for all $n\geq n_0$,
\[ \E\left[(\theta_{n+1}-\theta^*)^2\right] \leq C_1 \ell_n \log n, \]
for some $C_1$ that is independent of $n$. In particular, $C_1$ can be taken such that $C_1 > \frac{C_3}{1-\eta_2} \sup_{n\geq n_0} \frac{1 + \lambda^{-1}\log n + \lambda}{\log (n-1)}  $. The proof is completed.

\end{APPENDICES}

\end{document}